\documentclass[amsmath,amssymb,reprint,aps,prd,eqsecnum]{revtex4-2}

\usepackage{graphicx} 
\usepackage{mathrsfs}
\usepackage{caption}
\usepackage{subcaption}
\usepackage{xcolor}
\usepackage{hyperref}
\usepackage{comment}
\newcommand*\diff{\mathop{}\!\mathrm{d}}

\begin{document}

\title{Quantum-corrected anti-de Sitter space-time}

\author{Jacob C. Thompson}
\email{JThompson16@sheffield.ac.uk}

\affiliation{School of Mathematical and Physical Sciences,
The University of Sheffield,
Hicks Building,
Hounsfield Road,
Sheffield. S3 7RH United Kingdom}

\author{Elizabeth Winstanley}
\email{E.Winstanley@sheffield.ac.uk}

\affiliation{School of Mathematical and Physical Sciences,
The University of Sheffield,
Hicks Building,
Hounsfield Road,
Sheffield. S3 7RH United Kingdom}

\date{\today}

\begin{abstract}
We study the back-reaction of a quantum scalar field on anti-de Sitter (AdS) space-time.
The renormalized  expectation value of the stress-energy tensor operator for a massless, conformally-coupled quantum scalar field on global AdS space-time  in four space-time dimensions acts as a source term on the right-hand-side of the Einstein equations for the quantum-corrected metric.
We solve the quantum-corrected Einstein equations numerically and find deviations from pure AdS which 
increase as the temperature of the quantum scalar field state increases. 
We interpret these quantum-corrected metrics as asymptotically-AdS solitons, and study the mass of these solitons as a function of the temperature of the quantum scalar field. 
\end{abstract}

\maketitle

\section{Introduction}
\label{sec:intro}

The classical Einstein equations govern the interaction between space-time geometry and matter (we use units in which $8\pi G=c=1$):
\begin{equation}
\label{eq:classicalEinEqns}
     G_{\mu\nu} + \Lambda g_{\mu\nu} = T_{\mu\nu} ,
\end{equation}
where $G_{\mu \nu }$ is the Einstein tensor, $g_{\mu \nu }$ the metric tensor with signature $(-,+,+,+)$ and $\Lambda $ the cosmological constant.
The left-hand-side of (\ref{eq:classicalEinEqns}) describes the geometry of space-time.
On the right-hand-side, $T_{\mu \nu }$ is the classical stress-energy tensor (SET) of matter or classical fields other than gravity.
In a full theory of quantum gravity, both the left- and right-hand-sides of (\ref{eq:classicalEinEqns}) would be quantized, in other words, all matter fields, as well as the space-time geometry on which the matter propagates, would be quantum in nature.

Developing a full theory of quantum gravity is a rather ambitious challenge, and a more modest undertaking is to consider instead quantum field theory on curved space-time. 
In this paradigm, space-time (giving the quantities on the left-hand-side of (\ref{eq:classicalEinEqns})) continues to be classical but the matter (and fields other than gravity) is quantized.
This means that the classical SET on the right-hand-side of (\ref{eq:classicalEinEqns}) is replaced by the (renormalized) expectation value of a quantum stress-energy tensor (RSET) operator $\langle {\hat {T}}_{\mu \nu } \rangle $, giving the semiclassical Einstein equations (SCEE).

In four space-time dimensions, the SCEE take the form 
\begin{equation}
\label{eq:semiclassicalEinEqns}
     G_{\mu\nu} + \Lambda g_{\mu\nu} + \alpha H_{\mu \nu }^{(1)} + \beta H_{\mu \nu }^{(2)} = \langle {\hat {T}}_{\mu\nu} \rangle .
\end{equation}
As a result of the renormalization prescription for the RSET \cite{Christensen:1978yd,Birrell:1982ix,Decanini:2005eg}, the left-hand-side includes two additional, classical, conserved geometric tensors, $H^{(1,2)}_{\mu \nu }$, given by 
\begin{align}
H^{(1)}_{\mu \nu } & = 
\dfrac{1}{2}R^{2}g_{\mu \nu } -2RR_{\mu \nu }- 2g_{\mu \nu }\Box R + 2 \nabla _{\mu }\nabla _{\nu }R ,
\nonumber \\
H^{(2)}_{\mu \nu } & = 
\dfrac{1}{2} R^{\sigma \lambda }R_{\sigma \lambda } g_{\mu \nu } - \Box R_{\mu \nu } + \nabla _{\mu }\nabla _{\nu } R 
\nonumber \\  &  \qquad 
- \dfrac{1}{2}g_{\mu \nu }\Box R - 2R^{\sigma \lambda }R_{\sigma \mu \lambda \nu } ,
    \label{eq:H12}
\end{align}
where $R_{\mu \nu \lambda \sigma } $ is the Riemann tensor, $R_{\mu \nu }$ the Ricci tensor and $R$ the Ricci scalar, and two additional constants $\alpha $ and $\beta $. 
The tensors $H^{(1,2)}_{\mu \nu }$ contain four derivatives of the space-time metric.

A fully self-consistent solution of the SCEE would involve the RSET on the right-hand-side of \eqref{eq:semiclassicalEinEqns} being computed on the space-time background which gives the geometric quantities on the left-hand-side. 
This is, in general, a very challenging problem, particularly when the fourth-order terms $H^{(1,2)}_{\mu \nu }$ are included, and, furthermore, the equations (\ref{eq:semiclassicalEinEqns}) are known to possess runaway solutions \cite{Simon:1990ic,Parker:1993dk,Suen:1988uf}.
Nonetheless, rigorous work on solving the SCEE consistently can be found, for example, for a background Einstein static universe \cite{Sanders:2020osl}, cosmological space-times \cite{Pinamonti:2010is,Pinamonti:2013wya,Juarez-Aubry:2019jbw,Gottschalk:2022bte} and static space-times \cite{Juarez-Aubry:2020uim,Juarez-Aubry:2021abq}.

Given the difficulties inherent in solving the SCEE, there are various lines of attack in the literature to simplify the problem.
First, one can take a perturbative (in $\hbar $) approach (see, for example, \cite{Flanagan:1996gw,Taylor:2020urz}), in which the space-time metric takes the form
\begin{equation}
    g_{\mu \nu } = g_{\mu \nu }^{0} + \hbar g_{\mu \nu }^{(1)} + {\mathcal {O}}(\hbar ^{2}) ,
\end{equation}
where $g_{\mu \nu }^{0}$ is a given fixed background space-time metric and $g_{\mu \nu }^{(1)}$ is the first-order correction to the metric. 
Linearizing the SCEE \eqref{eq:semiclassicalEinEqns} (with the assumption that the constants $\alpha $ and $\beta $ are ${\mathcal {O}}(\hbar )$) yields a set of linear equations for $g_{\mu \nu }^{(1)}$ which are second-order in time derivatives \cite{Flanagan:1996gw,Taylor:2020urz}, and which involve, as a source term, the RSET computed on the fixed space-time metric $g_{\mu \nu }^{0}$. 
Even with the recent development of efficient methods of computing the RSET on black hole backgrounds (see, for example, \cite{Levi:2016quh,Levi:2016exv,Taylor:2022sly,Arrechea:2023fas}), solving these reduced equations is still difficult and often approximations to the RSET are employed to make progress (see, for example, \cite{Taylor:2020urz}).

An alternative approach is to employ an analytic approximation for the RSET which is valid on a general class of space-time metrics, so that the right-hand-side of \eqref{eq:semiclassicalEinEqns} can be written in closed form in terms of one or more unknown metric functions. 
The aim is then to solve \eqref{eq:semiclassicalEinEqns} exactly for these unknown metric functions.
For example, the Polyakov approximation \cite{Parentani:1994ij,Fabbri:2005mw} results from an $s$-wave approximation to the RSET on static, spherically symmetric space-times. 
This approximation has the advantage of involving only second-order derivatives of the metric, and hence it is possible to solve the SCEE \eqref{eq:semiclassicalEinEqns} under the simplifying assumption $\alpha = \beta = 0$. 
This strategy been used to find spherically symmetric semiclassical stars and black holes (see, for example, \cite{Carballo-Rubio:2017tlh,Arrechea:2019jgx,Arrechea:2021ldl,Arrechea:2021xkp,Arrechea:2023oax,Beltran-Palau:2022nec} for a selection of recent works using this approach).

In this paper we follow a third, even less ambitious, approach.
We fix a background space-time having metric $g^{0}_{\mu \nu }$ and find the RSET on this background. 
This RSET is then used as the source term in the SCEE \eqref{eq:semiclassicalEinEqns}, which are then solved nonperturbatively to give the quantum corrected (QC) metric $g_{\mu \nu }^{{\rm {QC}}}$ whose curvature tensors appear on the left-hand-side of \eqref{eq:semiclassicalEinEqns}.
This approach has been applied in three dimensions (where the additional curvature tensors $H^{(1,2)}_{\mu \nu }$ are absent) to find the QC BTZ black holes and naked singularities \cite{Casals:2016odj,Casals:2018ohr,Casals:2019jfo}, and also more recently to three-dimensional black holes in massive gravity \cite{Chernicoff:2024dll}.
In this paper, we will only be concerned with quantum states for which the constants $\alpha $ and $\beta $ in \eqref{eq:semiclassicalEinEqns} both vanish, so that the Einstein equations we will be solving can be written as
\begin{equation}
\label{eq:QCEinEqns}
     G_{\mu \nu}^{\rm {QC}} + \Lambda g_{\mu\nu}^{\rm {QC}}  =  \langle {\hat {T}}_{\mu\nu}[g^{0}] \rangle ,
\end{equation}
where the quantities on the left-hand-side involve the QC metric $g_{\mu \nu }^{\rm {QC}}$ while the RSET on the right-hand-side is computed on the fixed background metric $g_{\mu \nu }^{0}$.
The terms on the left-hand-side of (\ref{eq:QCEinEqns}) are second order in derivatives of the QC metric $g_{\mu \nu }^{\rm {QC}}$, which facilitates finding solutions.
This approach is often referred to as the ``back-reaction'' problem. 

In this paper we address the back-reaction problem for a simple toy model, taking the initial fixed background metric $g_{\mu \nu }^{0}$ to be that of four-dimensional global anti-de Sitter (AdS) space-time.
Since AdS is a maximally-symmetric space-time, this is the simplest nontrivial example of the back-reaction problem (the back-reaction on Minkowski space-time being trivial for a quantum field in the global vacuum state). 
Quantum field theory on four-dimensional global AdS was initiated many years ago \cite{Avis:1977yn,Allen:1986ty}  for a massless, conformally-coupled scalar field $\Phi$ satisfying the Klein-Gordon equation
\begin{equation}
    0 = \left( \nabla_\mu \nabla^\mu - \dfrac{1}{6}R\right)\Phi .
    \label{eq:scalar}
\end{equation}
There are two key features which render the back-reaction problem on pure AdS nontrivial. First, since AdS is not a globally hyperbolic space-time, it is necessary to apply boundary conditions to the quantum field at null infinity in order to have a well-defined theory (see, for example \cite{Avis:1977yn,Allen:1986ty,Ishibashi:2004wx,Dappiaggi:2018xvw,Morley:2020ayr}). 
The application of the simplest possible boundary conditions (namely Dirichlet and Neumann) yields a vacuum state which replicates the maximal symmetry of the underlying space-time \cite{Avis:1977yn,Allen:1986ty,Allen:1985wd,Kent:2014nya}, 
in which case the RSET in that state is a multiple of the AdS metric.
However maximal symmetry is broken in the vacuum state if more  general boundary conditions are considered, such as Robin boundary conditions \cite{Barroso:2019cwp,Morley:2020ayr,Morley:2023exv}. 
Maximal symmetry is also broken for thermal states \cite{Allen:1986ty}, even when the boundary conditions are such that the corresponding vacuum state is maximally symmetric. 
The back-reaction problem becomes nontrivial when the RSET on right-hand-side of the ``quantum-corrected'' Einstein equations (QCEE) (\ref{eq:QCEinEqns}) is evaluated in a nonmaximally-symmetric quantum state, since then it is no longer the case that the RSET is proportional to the space-time metric. 
Solving the back-reaction problem to find the QC metric  $g_{\mu \nu }^{\rm {QC}}$ in such a scenario is the focus of the present work. 

The outline of this paper is as follows.
We begin, in Sec.~\ref{sec:maxsym} by reviewing the (trivial) back-reaction problem for maximally-symmetric quantum states on maximally-symmetric space-times. 
In Sec.~\ref{sec:thermal} we review the RSET for a massless, conformally-coupled scalar field in a thermal state on pure AdS space-time, as computed in \cite{Allen:1986ty}. 
Our analysis of the QC metrics solving the QCEE (\ref{eq:QCEinEqns}) with the above RSET as a source on the right-hand-side begins in Sec.~\ref{sec:QCEE}, where we employ an ansatz for the QC metric which is particularly useful for comparing with the pure AdS metric. 
We derive the QCEE using this metric ansatz and outline our numerical procedure for solving these equations.
The results of this numerical work are presented in Sec.~\ref{sec:ABC}, where we consider the QC metrics resulting from applying  Dirichlet, Neumann and transparent boundary conditions to the quantum scalar field.
To aid the physical interpretation of these QC metrics, in Sec.~\ref{sec:solitons} we use an alternative ansatz for the QC metric, which enables us to interpret the QC metrics as soliton-like space-times which are asymptotically-AdS.  
We compute the masses of these solitons and investigate the dependence of the mass on the temperature of the quantum field.
Our conclusions are presented in Sec.~\ref{sec:conc}.

\section{QC maximally symmetric space-times}
\label{sec:maxsym}

We now review some simple examples of QC space-times, for which the RSET is known exactly in closed form for a massless, conformally-coupled, scalar field.
These examples involve background space-times which are maximally symmetric.
We first show that in this situation the constants $\alpha $ and $\beta $ in the SCEE \eqref{eq:semiclassicalEinEqns} can be set to zero if the RSET preserves the maximal symmetry of the background space-time.

\subsection{RSET for maximally symmetric states}
\label{sec:maxsymRSET}

Consider a background space-time which is maximally symmetric.
In this case the Riemann tensor is given in terms of the space-time metric:
\begin{equation}
    R_{\mu \nu \lambda \sigma } = {\mathcal {K}} \left(  g_{\mu \lambda }g_{\nu \sigma } - g_{\mu \sigma }g_{\nu \lambda } \right) 
\end{equation}
where ${\mathcal {K}}$ is a constant. 
In four space-time dimensions, the Ricci tensor and curvature scalar are then:
\begin{equation}
    R_{\mu \nu } = 3{\mathcal {K}} g_{\mu \nu }, \qquad R = 12{\mathcal {K}}.
\end{equation}
The geometric tensors $H_{\mu \nu }^{(1,2)}$  (\ref{eq:H12}) are therefore also proportional to the metric tensor $g_{\mu \nu }$ when evaluated on the background space-time.

Now consider a maximally-symmetric quantum state, for example, a global vacuum state (assuming such a state exists and is Hadamard).
In such a state the two-point function $G(x,x')$ is a function only of the world function $\sigma (x,x')$ for the separated space-time points $x$, $x'$ \cite{Allen:1985wd}.
Furthermore, the Hadamard parametrix $G_{\rm {H}}(x,x')$ (see, for example, \cite{Decanini:2005eg}), which must be subtracted from the two-point function as part of the renormalization process, is also a function only of $\sigma (x,x')$. 
The RSET is obtained from the regularized two-point function 
\begin{equation}
G_{\rm {R}}(x,x')=G(x,x')-G_{\rm {H}}(x,x')
\end{equation}
by applying a second-order differential operator ${\mathcal {T}}_{\mu \nu }$ 
and then taking the coincidence limit $x'\rightarrow x$. 
In other words, the RSET in a maximally-symmetric vacuum state $|0\rangle $ is
\begin{equation}
  \langle 0| {\hat {T}}_{\mu\nu} | 0\rangle  = \lim _{x'\rightarrow x} \left\{ {\mathcal {T}}_{\mu \nu } \left[ G_{\rm {R}}(x,x') \right]  \right\}   - g_{\mu \nu }v_{1}.
  \label{eq:maxsymT}
\end{equation}
For a massless, conformally-coupled scalar field, the operator is \cite{Decanini:2005eg}
\begin{align}
    {\mathcal {T}}_{\mu \nu } = & ~ \dfrac{2}{3}g_{\nu }{}^{\nu '}\nabla _{\mu }\nabla _{\nu '}
    + \dfrac{1}{6} g_{\mu \nu }g^{\rho \sigma '}\nabla _{\rho }\nabla _{\sigma '}
    \nonumber \\ & ~ 
    -\dfrac{1}{3} g_{\mu }{}^{\mu '}g_{\nu }{}^{\nu '} \nabla _{\mu '}\nabla _{\nu '}
    +\dfrac{1}{3} g_{\mu \nu }\nabla _{\rho }\nabla ^{\rho }
    + \dfrac{1}{6} G_{\mu \nu } ,
    \label{eq:mathcalT}
\end{align}
where $g_{\nu }{}^{\nu '}$ is the bivector of parallel transport and a prime $'$ denotes derivatives with respect to the space-time point $x'$.
The quantity $v_{1}$ in the final term in (\ref{eq:maxsymT}) is the coincidence limit of a state-independent, geometric biscalar which is necessary in order for the RSET to be conserved \cite{Decanini:2005eg}.
For a maximally-symmetric space-time, $v_{1}$ is a position-independent constant.

The singular terms ${\mathcal {T}}_{\mu \nu }[G_{\rm {H}}(x,x')]$ are purely geometric and give rise to a renormalization of the constants $8\pi G$ (which we have already set equal to unity), $\Lambda $, $\alpha $ and $\beta $ in the SCEE \eqref{eq:semiclassicalEinEqns}.
Since $G_{\rm {H}}(x,x')$ is a function only of $\sigma (x,x')$, using the form of the operator \eqref{eq:mathcalT} and the expansions of $g_{\nu }{}^{\nu '}\nabla _{\mu }\nabla _{\nu '}\sigma $ and $\nabla _{\mu }\nabla _{\nu }\sigma $ given in \cite{Decanini:2005gt}, it is straightforward to show that ${\mathcal {T}}_{\mu \nu }[G_{\rm {H}}(x,x')]$ is proportional to the metric tensor $g_{\mu \nu }$.
Therefore, it is sufficient to renormalize the cosmological constant $\Lambda $ in the SCEE \eqref{eq:semiclassicalEinEqns} and to set the constants $\alpha =0=\beta $. 

For the remainder of this paper, we will mostly be concerned with a QC metric in the situation where the original background space-time is maximally symmetric and possesses a maximally-symmetric vacuum state. 
We therefore set $\alpha = 0 =\beta $ in \eqref{eq:semiclassicalEinEqns} and from here on will study the solutions of the simplified SCEE
\begin{equation}
\label{eq:SCEE}
     G_{\mu}{}^{\nu} + \Lambda \delta _{\mu}{}^{\nu}  =  \langle {\hat {T}}_{\mu}{}^{\nu} \rangle ,
\end{equation}
where we have raised the second index for later convenience.
Suppose that the RSET for the maximally-symmetric vacuum state is given by 
\begin{equation}
    \langle 0 \lvert {\hat {T}}_{\mu}{}^{\nu} \rvert 0 \rangle =  {\mathcal {C}} \delta _{\mu }{}^{\nu },
\end{equation}
where ${\mathcal {C}}$ is a position-independent constant. 
Then, for any other quantum state $\rvert \psi  \rangle $ (\ref{eq:SCEE}) takes the form
\begin{equation}
    G_{\mu}{}^{\nu} + \Lambda \delta _{\mu}{}^{\nu}  = \langle \psi \lvert {\hat {T}}_{\mu}{}^{\nu}  \rvert \psi \rangle  - \langle 0 \lvert {\hat {T}}_{\mu}{}^{\nu} \rvert 0  \rangle  + {\mathcal {C}} \delta _{\mu }{}^{\nu },
\end{equation}
giving
\begin{equation}
\label{eq:QCEE}
    G_{\mu}{}^{\nu} + {\widetilde {\Lambda }} \delta _{\mu}{}^{\nu}  =   \langle \psi \lvert {\hat {T}}_{\mu}{}^{\nu}  \rvert \psi \rangle  - \langle 0 \lvert {\hat {T}}_{\mu}{}^{\nu} \rvert 0  \rangle ,
\end{equation}
where ${\widetilde {\Lambda }}= \Lambda -  {\mathcal {C}} $ is the renormalized cosmological constant. 
From now on we shall refer to (\ref{eq:QCEE}) as the ``quantum-corrected Einstein equations'' (QCEE).

\subsection{Examples}
\label{sec:examples}

Before we study solutions of (\ref{eq:QCEE}) for a background global AdS space-time, we review some simple examples of maximally-symmetric states on maximally symmetric space-times. 

\subsubsection{Minkowski space-time}
\label{sec:Minkowski}

The simplest possible maximally-symmetric space-time background is Minkowski space-time, whose metric in static, spherically symmetric coordinates $(t,r,\theta, \varphi )$ takes the form
\begin{equation}
    \diff{s}^{2} = -\diff{t}^{2} + \diff{r}^{2} + r^{2} \diff{\Omega } ^{2},
\end{equation} 
where 
\begin{equation}
\diff{\Omega } ^{2}= \diff{\theta } ^{2} + \sin ^{2}\theta \, \diff{\varphi }^{2}
    \label{eq:s2metric}
\end{equation}
is the metric on the two-sphere.
In the global Minkowski vacuum $|0_{{\rm {M}}} \rangle $, the RSET $\langle 0_{{\rm {M}}} | {\hat {T}}_{\mu\nu}  |0_{{\rm {M}}} \rangle \equiv 0$.
In this case there are no quantum corrections to the space-time metric and the back-reaction problem is trivial. 

\subsubsection{De Sitter space-time}
\label{sec:dS}

De Sitter (dS) space-time is maximally-symmetric space-time with positive scalar curvature. 
In global coordinates $(\tau, \rho, \theta , \varphi )$ the metric can be written as
\begin{equation}
    \diff{s}^{2} = a^{-2} \sec ^{2}\tau \left[ -\diff{\tau }^{2} + \diff{\rho } ^{2} + \sin ^{2}\rho \, \diff{\Omega }^{2} \right] ,
\end{equation}
where the cosmological constant is $\Lambda = 3a^{2}>0$ and $a$ is an inverse length scale.
The preferred quantum state on dS is the Bunch-Davies state \cite{Bunch:1978yq,Allen:1985ux}, which is the unique maximally-symmetric Hadamard state for a massless, conformally-coupled, scalar field. 
The RSET in this state is (setting $\hbar =1$ from henceforth) \cite{Bunch:1978yq,Page:1982fm,Tadaki:1988cn}
\begin{equation}
    \langle 0_{{\rm {BD}}} | {\hat {T}}_{\mu}{}^{\nu} | 0 _{{\rm {BD}}} \rangle 
     =  \dfrac{a^{2}}{960\pi ^{2}} \delta _{\mu }{}^{\nu } .
     \label{eq:RSETBD}
\end{equation}
In this case the QCEE \eqref{eq:QCEE} give the QC metric to be dS space-time with a renormalized cosmological constant:
\begin{equation}
    \Lambda \rightarrow {\widetilde {\Lambda }}_{\rm{dS}} = \Lambda -  \dfrac{a^{2}}{960\pi ^{2}}  .
    \label{eq:LambdadS}
\end{equation}
Solutions of the SCEE of dS form have been studied extensively in the literature: see, for example,
\cite{Dappiaggi:2008mm,Juarez-Aubry:2019jbw,Gottschalk:2022bte}.

\subsubsection{Anti-de Sitter space-time}
\label{sec:adS}

Anti-de Sitter (AdS), like dS, is a maximally-symmetric space-time, the difference being that AdS has  constant negative scalar curvature. In global coordinates $(\tau, \rho, \theta , \varphi )$ the metric is written as
\begin{equation}
\label{eq:pureAdSmetric}
    \diff{s}^{2} = a^{-2}\sec^2\rho \left[ -\diff{\tau}^2 + \diff{\rho}^2 +\sin^2\rho\diff{\Omega}^2 \right] ,
\end{equation}
where $a$ is the inverse AdS length scale, related to the cosmological constant by $\Lambda = -3a^{2} <0$. 
The coordinate ranges are $\tau\in (-\pi,\pi]$, $\rho\in [0,\pi/2)$, $\theta\in [0,\pi]$ and $\phi\in [0,2\pi]$. 

In order to have a well-defined quantum field theory on AdS, it is necessary to impose boundary conditions on the massless, conformally-coupled, scalar field \cite{Avis:1977yn,Ishibashi:2004wx,Dappiaggi:2018xvw,Morley:2020ayr}.
In this work we consider Dirichlet, Neumann and transparent boundary conditions \cite{Avis:1977yn}. 
In this case the vacuum state for the scalar field preserves the underlying maximal symmetry of the AdS space-time \cite{Avis:1977yn,Kent:2014nya,Morley:2020ayr} and the RSET in the vacuum state, $\langle 0 \lvert {\hat {T}}_{\mu}{}^{\nu} \rvert 0 \rangle$, is proportional to the metric~\cite{Allen:1986ty}
\begin{equation}
    \langle 0 \rvert {\hat {T}}_{\mu}{}^{\nu} \rvert 0 \rangle = -\dfrac{a^{2}}{960\pi ^{2}}\delta_{\mu}{}^{\nu}.
    \label{eq:adSvev}
\end{equation}
As in the de Sitter case, the QCEE give the QC metric in this case to be AdS space-time with a renormalized cosmological constant:
\begin{equation}
    \Lambda \rightarrow {\widetilde {\Lambda }}_{\rm{adS}} = \Lambda +  \dfrac{a^{2}}{960\pi ^{2} } .
    \label{eq:adSLren}
\end{equation}
The corrections to the cosmological constant due to the vacuum RSET are very small in both the dS (\ref{eq:LambdadS}) and AdS (\ref{eq:adSLren}) cases and are such that the magnitude of the cosmological constant is reduced.

\section{RSET for thermal states on AdS}
\label{sec:thermal}

In the previous section, we have considered only maximally-symmetric states, namely the vacuum state on Minkowski space-time, the Bunch-Davies state on dS, and the vacuum state on global AdS with either transparent, Dirichlet or Neumann boundary conditions. 
We now switch our focus to nonmaximally-symmetric states, in particular thermal states for a massless, conformally-coupled, scalar field.
The global thermal state on Minkowski space-time is spatially homogeneous and isotropic, so the solution of the QCEE (\ref{eq:QCEE}) is not particularly interesting in this case.
Due to the presence of the cosmological horizon, there is a preferred temperature on the static patch of dS. The Bunch-Davies state corresponds, on the static patch of dS, to a state at this preferred temperature, and therefore the solution of the QCEE (\ref{eq:QCEE}) is trivial in this case. 
For the remainder of this paper we therefore focus on thermal states on AdS.
In this section we review the thermal RSET as computed in \cite{Allen:1986ty}. 
Two aspects make this interesting for our purposes: first, AdS has no preferred temperature so we can consider the effect of varying the temperature; and second, the thermal RSET is not maximally symmetric.

The quantum scalar field ${\hat {\Phi }}$ is taken to be in a thermal state at inverse temperature $\beta $ and for the remainder of this paper we set $k_{{\rm {B}}}=1$.
The difference in the RSET between the thermal $\rvert \beta \rangle $ and vacuum $\rvert 0 \rangle $ states is \cite{Allen:1986ty} (with a minor typographical error corrected): 
\begin{subequations}
\label{eq:RSET}
\begin{align}
{\widetilde {T}}_{\mu }{}^{\nu }
& = 
    \langle \beta \lvert  {\hat {T}}_{\mu}{}^{\nu} \rvert \beta  \rangle 
    -  \langle 0 \lvert  {\hat {T}}_{\mu}{}^{\nu} \rvert 0  \rangle 
    \nonumber \\ 
&    = 
    \dfrac{a^{2}}{8\pi ^{2}}
    \left\{  {\mathcal {F}}_{1}(\rho ) \delta_{\mu}{}^{\nu} + {\mathcal {F}}_{2}(\rho ) \tau _{\mu} \tau^{\nu } + {\mathcal {F}}_{3}(\rho)\rho_{\mu }\rho^{\nu }
    \right\}
    \label{eq:RSETF}
\end{align}
where
\begin{equation}
\label{eq:RSETvecfields}
    \tau _\bullet = (a^{-1}\sec\rho,0,0,0), \qquad \rho_\bullet = (0,a^{-1}\sec\rho,0,0),
\end{equation}
are unit length timelike and radial vectors, respectively.
The functions ${\mathcal {F}}_{1}(\rho )$, ${\mathcal {F}}_{2}(\rho )$ and ${\mathcal {F}}_{3}(\rho )$ are given by
\begin{widetext}
\begin{align}
{\mathcal {F}}_{1}(\rho ) = & ~ \dfrac{4}{3}\cos^4 \rho \, f_3(a\beta)
+\lambda  \cot \rho  \left[ -\dfrac{1}{6}\csc^2 \rho\cos ( 2\rho)S_0(a\beta,\rho)
+\dfrac{1}{3}\cot \rho \, C_1(a\beta,\rho) 
+\dfrac{2}{3}\cos^2 \rho \,  S_2(a\beta,\rho) \right] ,
\nonumber \\
{\mathcal {F}}_{2}(\rho ) = & ~ \dfrac{16}{3}\cos^4 \rho \, f_3(a\beta) 
+\lambda  \cot \rho  \left[ \dfrac{1}{6}\left(3-\cot^2 \rho \right)S_0(a\beta,\rho)
+\cot \rho\left(1-\dfrac{2}{3}\cos^2 \rho \right)C_1(a\beta,\rho) 
\right. \nonumber \\ & \qquad  \left. 
+2\cos^2 \rho \, S_2(a\beta,\rho)\right] ,
\nonumber \\
{\mathcal {F}}_{3}(\rho ) = &  ~
\lambda  \cot \rho
\left[ \dfrac{1}{6}\left(3\csc^2 \rho -4\right)S_0(a\beta,\rho)
+\cot \rho \left(\dfrac{2}{3}\sin^2 \rho -1\right)C_1(a\beta,\rho)-\dfrac{2}{3}\cos^2 \rho \, S_2(a\beta,\rho) \right] ,
\label{eq:H}
\end{align}
\end{widetext}
where we have defined
\begin{align}
    f_m(x) &= \sum_{n=1}^{\infty} n^m \left(e^{nx}-1\right)^{-1}, \label{eq:InfSumfAFG}\\
     \widetilde{f}_m(x) &= \sum_{n=1}^{\infty} n^m (-1)^n \left(e^{nx}-1\right)^{-1}, \label{eq:InfSumfaltAFG}\\
    S_m(x,\rho) &= \sum_{n=1}^{\infty} n^m (-1)^n \left(e^{nx}-1\right)^{-1}\sin (2n\rho), \label{eq:InfSumSinAFG} \\
    C_m(x,\rho) &= \sum_{n=1}^{\infty} n^m (-1)^n \left(e^{nx}-1\right)^{-1}\cos (2n\rho), \label{eq:InfSumCosAFG}
\end{align}
and the second quantity ${\widetilde {f}}_{m}(x)$ has been defined for later convenience.
In (\ref{eq:H}), the constant $\lambda $ depends on the boundary conditions imposed:
\begin{equation}
\lambda = \begin{cases}
    1 & {\mbox {Dirichlet boundary conditions,}}
    \\
    -1 & {\mbox {Neumann boundary conditions,}}
    \\ 
    0 & {\mbox {transparent boundary conditions.}}
\end{cases}   
\label{eq:lambda}
\end{equation}
\end{subequations}
When $\lambda = 0$ and we consider transparent boundary conditions, we have ${\mathcal {F}}_{3}(\rho )=0$ and the functions ${\mathcal {F}}_{1,2}(\rho )$ simplify considerably:
\begin{align}
{\mathcal {F}}_{1}(\rho ) = & ~ \dfrac{4}{3}\cos^4 \rho \, f_3(a\beta) ,
\nonumber \\
{\mathcal {F}}_{2}(\rho ) = & ~ \dfrac{16}{3}\cos^4 \rho \, f_3(a\beta).
\end{align}
In the limit $\beta \rightarrow \infty $,  the functions $f_{m}(a\beta )$, $S_{m}(a\beta , \rho )$ and $C_{m}(a\beta ,\rho)$  all tend to zero and the difference in  RSET expectation values \eqref{eq:RSET} vanishes.

\begin{figure}
\centering
\begin{subfigure}{\columnwidth}
    \includegraphics[width=\columnwidth]{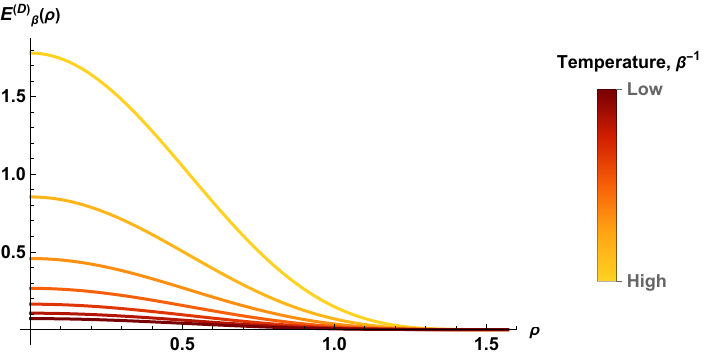}
    \caption{Energy density}
    \label{fig:EnDenPlotD}
\end{subfigure}
\begin{subfigure}{\columnwidth}
    \includegraphics[width=\columnwidth]{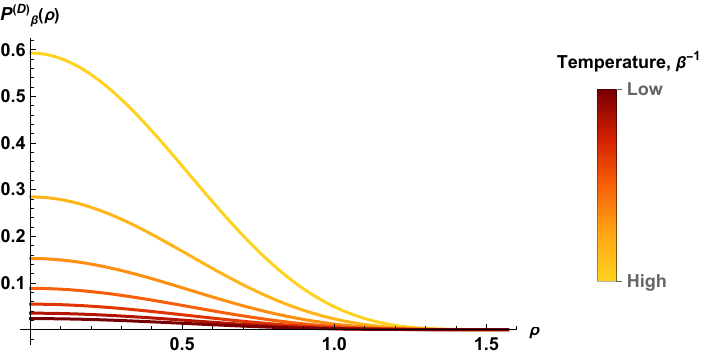}
    \caption{Pressure}
    \label{fig:PresPlotD}
\end{subfigure}
\begin{subfigure}{\columnwidth}
    \includegraphics[width=\columnwidth]{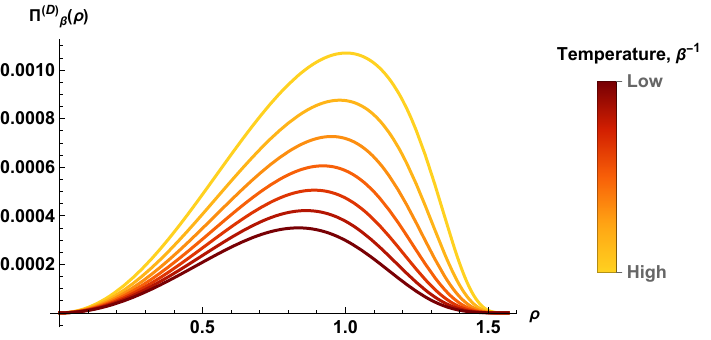}
    \caption{Pressure deviator}
    \label{fig:PresDevPlotD}
\end{subfigure}    
\caption{Energy density $E_{\beta }^{(\rm {D})}(\rho )$, pressure $P_{\beta }^{(\rm {D})}(\rho )$ and pressure deviator $\Pi _{\beta }^{(\rm {D})}(\rho )$ (\ref{eq:RKTquantities}) for a massless, conformally-coupled, scalar field on global AdS with Dirichlet boundary conditions applied, for varying inverse temperature $\beta \in (\tfrac{\pi}{12}, \tfrac{11\pi }{24})$ and fixed inverse AdS length scale $a=1$.}
\label{fig:RSETplotsD}
\end{figure}

\begin{figure}
    \centering
    \includegraphics[width=\columnwidth]{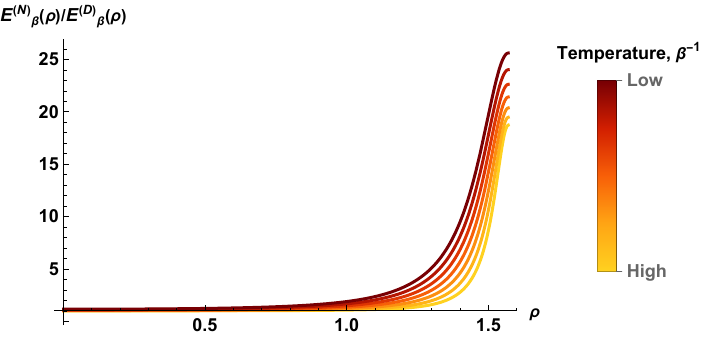}
    \caption{Ratio of the energy density $E_{\beta }^{(\rm N)}(\rho) $ with Neumann boundary conditions applied to the energy density $E_{\beta }^{(\rm D)}(\rho )$ with Dirichlet boundary conditions applied. The inverse AdS length scale is fixed to be $a=1$ and the inverse temperature $\beta \in (\tfrac{\pi}{12}, \tfrac{11\pi }{24})$.}
    \label{fig:EnergyRatio}
\end{figure}

From the form of the RSET (\ref{eq:RSET}), we have ${\widetilde {T}}_{\theta }{}^{\theta } = {\widetilde {T}}_{\varphi }{}^{\varphi }$ and therefore it suffices to consider the three components ${\widetilde {T}}_{\tau }{}^{\tau }$, ${\widetilde {T}}_{\rho }{}^{\rho }$ and ${\widetilde {T}}_{\theta }{}^{\theta }$. If we define
\begin{subequations}
\label{eq:RKTquantities}
\begin{align}
    E_{\beta}(\rho) &=  - \widetilde{T}_{\tau}{}^{\tau}, \\
    P_{\beta}(\rho)&= \dfrac{1}{3} \left[  \widetilde{T}_{\rho}{}^{\rho} + 2\widetilde{T}_{\theta}{}^{\theta} \right],
\\
    \Pi_{\beta}(\rho) &   = \dfrac{2}{3}\left[  \widetilde{T}_{\rho}{}^{\rho} - \widetilde{T}_{\theta}{}^{\theta} \right] ,
\end{align}   
these are the (local) energy density, pressure and pressure deviator respectively~\cite{Ambrus:2018olh}. 
\end{subequations}

In Fig.~\ref{fig:RSETplotsD}, we show the energy density, pressure and pressure deviator when Dirichlet boundary conditions are applied. 
All three quantities (energy density, pressure and pressure deviator) tend toward zero as the temperature is decreased. This corresponds to the thermal RSET tending towards the vacuum expectation value of the RSET in the low-temperature limit, as expected. 
The energy density and pressure for Neumann or transparent boundary conditions are qualitatively extremely similar to those shown in Fig.~\ref{fig:RSETplotsD}.

In Fig.~\ref{fig:EnergyRatio}, we show the ratio of the energy density $E_{\beta }^{(\rm N)}(\rho )$ when Neumann boundary conditions are applied to the scalar field to the energy density $E_{\beta }^{(\rm D)}(\rho )$ when Dirichlet boundary conditions are applied. 
This ratio is appreciably different from unity close to the boundary.
While both $E_{\beta }^{(\rm N)}(\rho )$ and $E_{\beta }^{(\rm D)}(\rho )$ tend to zero as $\rho \rightarrow \tfrac{\pi }{2}$, we see from Fig.~\ref{fig:EnergyRatio} that the energy density for Neumann boundary conditions is tending to zero more slowly than the energy density for Dirichlet boundary conditions.
It is notable from Fig.~\ref{fig:EnergyRatio} that the ratio $E_{\beta }^{(\rm N)}(\rho )/E_{\beta }^{(\rm D)}(\rho )$ increases as the temperature decreases. 

Similar behaviour is observed for the ratio of the energy density $E_{\beta }^{(\rm T)}(\rho )$ for transparent boundary conditions to the Dirichlet energy density $E_{\beta }^{(\rm D)}(\rho )$, except that $E_{\beta }^{(\rm T)}(\rho )/E_{\beta }^{(\rm D)}(\rho )$ is approximately half $E_{\beta }^{(\rm N)}(\rho )/E_{\beta }^{(\rm D)}(\rho )$.
The corresponding pressure ratios $P_{\beta }^{(\rm N)}(\rho )/P_{\beta }^{(\rm D)}(\rho )$ and $P_{\beta }^{(\rm T)}(\rho )/P_{\beta }^{(\rm D)}(\rho )$ have similar properties.
The pressure deviator for the case of Neumann boundary conditions  is exactly minus the pressure deviator in the case of Dirichlet boundary conditions, that is, $\Pi_{\beta}^{(\rm D)}(\rho) = - \Pi_{\beta}^{(\rm N)}(\rho)$.
In the case of transparent boundary conditions the pressure deviator is identically zero. 

Unlike the RSET in the vacuum state, we see from Fig.~\ref{fig:RSETplotsD} that the thermal RSET is not maximally symmetric.
A maximally-symmetric state has constant energy density which equals the pressure, and vanishing pressure deviator.
The energy density and pressure distributions shown in Fig.~\ref{fig:RSETplotsD} tend to `clump' towards the origin at $\rho =0$  and tend to zero as the space-time boundary at $\rho = \pi /2$ is approached \cite{Ambrus:2018olh}.
In contrast to the situations discussed in Sec.~\ref{sec:examples}, we therefore expect that the QC metric resulting from the thermal RSET will not be maximally symmetric.
Finding this QC metric is the focus of the remainder of the paper.

\section{QC Einstein equations}
\label{sec:QCEE}

In this section we derive the explicit form of the QCEE (\ref{eq:QCEE}) governing QC AdS space-times and outline our numerical method for solving these equations.

\subsection{QCEE for QC AdS}
\label{sec:SCEErho}

We start with our ansatz for the QC metric, which takes the form
\begin{equation}
\label{eq:QCadSansatz}
    \diff{{s}}_{\rm {QC}}^2 = -A(\rho)\diff{\tau}^2+B(\rho)\diff{\rho}^2+C(\rho)\diff{\Omega}^2
\end{equation}
where $A(\rho )$, $B(\rho )$ and $C(\rho )$ are functions of $\rho$, to be determined by the QCEE (\ref{eq:QCEE}):
\begin{equation}
G_{\mu}{}^{\nu} + {\widetilde {\Lambda }} \delta _{\mu}{}^{\nu}  =  {\widetilde  {T}}_{\mu}{}^{\nu}, \label{eq:QCEEAdS}
\end{equation}
where ${\widetilde {T}}_{\mu}{}^{\nu }$ is the difference between the thermal and vacuum RSET (\ref{eq:RSET}) and ${\widetilde {\Lambda }}={\widetilde {\Lambda }}_{\rm {AdS}}$ is the renormalized cosmological constant given by (\ref{eq:adSLren}).
With the metric ansatz (\ref{eq:QCadSansatz}), the QCEE (\ref{eq:QCEEAdS}) take the form
\begin{subequations}
\label{eq:semiclassEinThreeEqns}
\begin{align}
   {\widetilde {T}}_{\tau }{}^{\tau }     = & ~ \dfrac{4B^2C({\widetilde {\Lambda }} C-1) - 2B'CC'-B(C'^2-4CC'')}{4B^2C^2} , 
 \label{eq:semiclassEintt}
 \\
   {\widetilde {T}}_{\rho}{}^{\rho} = & ~
    \dfrac{2A'CC'+A(4BC[{\widetilde {\Lambda }} C-1]+C'^2)}{4ABC^2} , 
    \label{eq:semiclassEinrhorho} 
    \\
       {\widetilde {T}}_{\theta}{}^{\theta} = & ~
    -\dfrac{1}{4A^2B^2C^2}\left[A'^2BC^2
    \right. \notag \\ & \left. -AC(A'BC'
    +C\{ 2A''B-A'B'\} )
    \right.\notag\\
    &  -\left.
    A^2(4{\widetilde {\Lambda}} B^2C^2-B'CC'-B\{ C'^2-2CC''\} )\right], 
    \label{eq:semiclassEinthetatheta}
\end{align}
\end{subequations}
where a prime denotes differentiation with respect to $\rho $, and the $\varphi$ component has been omitted as ${\widetilde {T}}_{\theta}{}^{\theta} =  {\widetilde {T}}_{\varphi}{}^{\varphi}$. 

Using (\ref{eq:semiclassEintt}, \ref{eq:semiclassEinrhorho}), we find that \eqref{eq:semiclassEinthetatheta} can be written as
\begin{equation} 
\label{eq:equationrelations}
    \left( {\widetilde {T}}_{\theta}{}^{\theta}
    -{\widetilde {T}}_{\rho}{}^{\rho}\right)
    -\dfrac{C}{C'}{\widetilde {T}}_{\rho}{}^{\rho }{}'
    +\dfrac{A'C}{2AC'}\left({\widetilde {T}}_{\tau }^{\tau }-{\widetilde {T}}_{\rho}{}^{\rho}\right) = 0.
\end{equation}
Taking the covariant derivative of the RSET gives
\begin{multline}
    \nabla_{\mu}{\widetilde {T}}_{\rho}{}^{\mu} = \\ \dfrac{1}{2B}\left[
    \dfrac{2C'}{C}
    \left({\widetilde {T}}_{\rho}{}^{\rho}
    -{\widetilde {T}}_{\theta}{}^{\theta}\right)
    +2{\widetilde {T}}_{\rho}{}^{\rho}{}'
    +\dfrac{A'}{A}\left(
    {\widetilde {T}}_{\rho}{}^{\rho}
    -{\widetilde {T}}_{\tau }{}^{\tau }\right)\right] ,
\end{multline}
which is precisely \eqref{eq:equationrelations} multiplied by $-\tfrac{C'}{BC}$. 
Hence, provided the QCEE~(\ref{eq:semiclassEintt}--\ref{eq:semiclassEinthetatheta}) hold, conservation of the RSET follows, as is expected. 
Thus the QCEE and RSET conservation give three independent equations for three unknown functions $A(\rho )$, $B(\rho )$ and $C(\rho )$. 

To simplify these, we first rearrange (\ref{eq:semiclassEintt}, \ref{eq:semiclassEinrhorho}, \ref{eq:equationrelations}) to give
\begin{subequations}
\label{eq:redEin123}
\begin{align}
    0 &= {\widetilde {\Lambda }} -  {\widetilde {T}}_{\tau}{}^{\tau } 
    - \dfrac{C'^2}{4BC^2}+\dfrac{2BC''-2B^2-B'C'}{2B^2C} ,
    \label{eq:redEin1} 
    \\
    0 &= {\widetilde {\Lambda }} 
    -   {\widetilde {T}}_{\rho}{}^{\rho} 
    + \dfrac{C'^2}{4BC^2}+\dfrac{A'C'-2AB}{2ABC} ,\label{eq:redEin2}
    \\
    0 &= A'C( {\widetilde {T}}_{\rho}{}^{\rho}
    -{\widetilde {T}}_{\tau }{}^{\tau }) +2A(C{\widetilde {T}}_{\rho}{}^{\rho}{})'
    -2AC' {\widetilde {T}}_{\theta}{}^{\theta } . 
    \label{eq:redEin3}
\end{align}
\end{subequations}
Eq.~\eqref{eq:redEin3} can be rearranged to give $A'(\rho )/A(\rho )$ in terms of $C(\rho )$ and RSET components:
\begin{equation}
    \dfrac{A'}{A} = 2\dfrac{(C{\widetilde {T}}_{\rho}{}^{\rho}{})'-C'{\widetilde {T}}_{\theta}{}^{\theta } }{C({\widetilde {T}}_{\tau }{}^{\tau }-{\widetilde {T}}_{\rho}{}^{\rho})}. 
    \label{eq:Aeqn}
\end{equation}
We use this equation to eliminate $A'/A$ from \eqref{eq:redEin2}, to give 
\begin{multline} 
\label{eq:redEin2a}
    0 = {\widetilde {\Lambda }} 
    -   {\widetilde {T}}_{\rho}{}^{\rho}
    - \dfrac{1}{C} + \dfrac{C'^2}{4BC^2}\\-\dfrac{C'(C{\widetilde {T}}_{\rho}{}^{\rho}{}'
    +C'{\widetilde {T}}_{\rho}{}^{\rho}
    -C'{\widetilde {T}}_{\theta}{}^{ \theta})}{BC^2({\widetilde {T}}_{\rho}{}^{\rho}
    -{\widetilde {T}}_{\tau }{}^{\tau })}.
\end{multline}
We can now rearrange (\ref{eq:redEin2a}) to give an expression for $B(\rho )$ in terms of $C(\rho ) $ (and its derivatives) and RSET components (and derivatives thereof):
\begin{equation}
   B =\dfrac{C'^2({\widetilde {T}}_{\rho}{}^{\rho}
    -{\widetilde {T}}_{\tau }{}^{\tau })-4C'(C{\widetilde {T}}_{\rho}{}^{\rho}{}'
    +C'{\widetilde {T}}_{\rho}{}^{\rho}
    -C'{\widetilde {T}}_{\theta}{}^{ \theta})}{4C({\widetilde {T}}_{\rho}{}^{\rho}
    -{\widetilde {T}}_{\tau }{}^{\tau })(1-C{\widetilde {\Lambda }} 
    +  C {\widetilde {T}}_{\rho}{}^{\rho})} .
    \label{eq:Beqn}
\end{equation}
We then substitute for $B(\rho )$ and $B'(\rho )$ from (\ref{eq:Beqn}) into~\eqref{eq:redEin1} to yield a single equation for $C$: 
\begin{subequations}
\label{eq:singleCeqnFull}
\begin{align}  
 \notag  0 & = \dfrac{1}{CC'}\left( 2CC'' + C'^2 \right) \dfrac{\left(  {\widetilde {T}}_{\rho}{}^{\rho} - {\widetilde {T}}_{\tau }{}^{\tau } \right)\Xi_\rho}{\Upsilon} \\
   &
 \quad 
 +2\dfrac{\diff}{\diff{\rho}}\left(\dfrac{\left(  {\widetilde {T}}_{\rho}{}^{\rho} - {\widetilde {T}}_{\tau }{}^{\tau } \right)\Xi_\rho}{\Upsilon}\right) + \dfrac{\Xi_\tau}{C} , 
   \label{eq:singleCeqn}  
\end{align}
where we have defined
\begin{equation}
    \Upsilon = C'\left(  {\widetilde {T}}_{\tau }{}^{\tau } 
    +3 {\widetilde {T}}_{\rho}{}^{\rho}     
-4{\widetilde {T}}_{\theta}{}^{\theta} \right)
+4 C
   {\widetilde {T}}_{\rho}{}^{\rho} {}'
   \label{eq:Upsilon}
\end{equation}
and
\begin{equation}
    \Xi_\bullet = C\left( {\widetilde {\Lambda }} -  {\widetilde {T}}_{\bullet}{}^{\bullet} \right) -1 .
    \label{eq:Xi}
\end{equation}
\end{subequations}
We have therefore reduced the QCEE to a single, second order, nonlinear ODE (ordinary differential equation) for the metric function $C(\rho )$. 
Once $C(\rho )$ is known, we can then use (\ref{eq:Beqn}) to give the metric function $B(\rho )$ and also solve the first order ODE (\ref{eq:Aeqn}) to give the remaining metric function $A(\rho )$.
To solve these ODEs, we require boundary conditions on the metric functions, which are described next.

\subsection{Solving the QC Einstein equations}
\label{sec:method}

The QCEE (\ref{eq:Aeqn}, \ref{eq:Beqn}, \ref{eq:singleCeqnFull}) are sufficiently complicated that proceeding numerically is required to find the QC metric functions $A(\rho )$, $B(\rho )$ and $C(\rho )$. 
Our assumptions about the form of $C(\rho )$ (see (\ref{eq:ABCseries}) below) mean that these equations are singular at $\rho =0$. 
We will start our numerical integration at $\rho =\epsilon $ for some positive constant $\epsilon \ll 1$ and integrate for increasing $\rho $ towards the space-time boundary at $\rho = \tfrac{\pi }{2}$. 
For this procedure, we require boundary conditions on the QC metric functions at $\rho = \epsilon $.

We first perform a Taylor series expansion near $\rho =0$ of the three functions ${\mathcal {F}}_{i}$, $i=1,2,3$ (\ref{eq:H}), which take the form:
\begin{widetext}
\begin{align}
    {\mathcal {F}}_{1}(\rho )= & ~
    \dfrac{4}{9}\left[ \lambda \widetilde{f}_{1}(a\beta ) + 3 f_{3}(a\beta) + 2\lambda  \widetilde{f}_{3}(a\beta ) \right]  
    -\dfrac{8}{45}\left[  \lambda \widetilde{f}_{1}(a\beta ) 
    +15 f_{3}(a\beta ) + 10\lambda \widetilde{f}_{3}(a\beta ) +4\lambda \widetilde{f}_{5}(a\beta ) \right] \rho ^{2} + 
   {\mathcal {O}}(\rho ^{4}) ,
   \nonumber \\ 
   {\mathcal {F}}_{2}(\rho )= & ~
    \dfrac{16}{9}\left[ \lambda \widetilde{f}_{1}(a\beta ) +  3 f_{3}(a\beta )+ 2\lambda  \widetilde{f}_{3}(a\beta ) \right] 
    -\dfrac{16}{45} \left[ 3\lambda \widetilde{f}_{1}(a\beta ) +30 f_{3}(a\beta ) + 20\lambda \widetilde{f}_{3}(a\beta ) + 7\lambda \widetilde{f}_{5}(a\beta ) \right] \rho ^{2} + 
   {\mathcal {O}}(\rho ^{4}) ,
   \nonumber \\ 
   {\mathcal {F}}_{3}(\rho ) = & ~ 
    -\dfrac{16\lambda }{45}\left[ \widetilde{f}_{1}(a\beta ) - \widetilde{f}_{5}(a\beta ) \right] \rho ^{2} + 
   {\mathcal {O}}(\rho ^{4}) , 
   \label{eq:Fseries}
    \end{align}
\end{widetext}
where $f_{m}(a\beta )$ and $\widetilde{f}_{m}(a\beta)$ are given in (\ref{eq:InfSumfAFG}--\ref{eq:InfSumfaltAFG}). 
Due to the form of the functions in (\ref{eq:H}), the above series expansions for small $\rho  $ involve only even powers of $\rho $.
Substituting the series (\ref{eq:Fseries}) into the expression for the RSET (\ref{eq:RSETF}), the RSET components have the following series expansions for small $\rho $:
\begin{subequations}
\label{eq:RSETseries}
\begin{align}
    {\widetilde {T}}_{\tau} {}^{\tau } (\rho ) & = ~
    T_{\tau 0} + T_{\tau 2 } \rho ^{2} + {\mathcal {O}}(\rho ^{4}) ,
    \nonumber \\ 
        {\widetilde {T}}_{\rho } {}^{\rho } (\rho ) & = ~
        T_{\rho 0} + T_{\rho 2 } \rho ^{2} + {\mathcal {O}}(\rho ^{4}) ,
    \nonumber \\ 
    {\widetilde {T}}_{\theta  } {}^{\theta } (\rho ) & = ~
    T_{\theta 0} + T_{\theta  2 } \rho ^{2} + {\mathcal {O}}(\rho ^{4}) ,
\end{align}
where
\begin{align}
    T_{\tau 0 } & = ~ -\dfrac{a^{4}}{6\pi ^{2}}\left[ \lambda \widetilde{f}_{1}(a\beta ) + 3 f_{3}(a\beta ) + 2\lambda  \widetilde{f}_{3}(a\beta ) \right]  ,
    \nonumber \\
    T_{\tau 2 } & = ~ \dfrac{a^{4}}{9\pi ^{2}} 
    \left[  \lambda\widetilde{f}_{1}(a\beta ) 
    +9 f_{3}(a\beta ) + 6\lambda\widetilde{f}_{3}(a\beta ) \right. \nonumber\\ & \qquad \left. + 2\lambda \widetilde{f}_{5}(a\beta )
    \right],
        \nonumber \\
    T_{\rho 0} & = ~ \dfrac{a^{4}}{18\pi ^{2}}\left[ \lambda \widetilde{f}_{1}(a\beta ) + 3f_{3}(a\beta ) + 2\lambda \widetilde{f}_{3}(a\beta )\right]  ,
        \nonumber \\
    T_{\rho 2} & = ~ -\dfrac{a^{4}}{45\pi ^{2}}
\left[  3\lambda \widetilde{f}_{1}(a\beta ) 
    +15  f_{3}(a\beta ) + 10\lambda\widetilde{f}_{3}(a\beta )
    \right. \nonumber\\ & \qquad \left .+ 2\lambda \widetilde{f}_{5}(a\beta )\right],
        \nonumber \\
        T_{\theta 0} & = ~ T_{\rho 0} = \dfrac{a^{4}}{18\pi ^{2}}\left[ \lambda \widetilde{f}_{1}(a\beta ) + 3f_{3}(a\beta ) + 2\lambda \widetilde{f}_{3}(a\beta )\right] ,
            \nonumber \\
            T_{\theta 2} & = ~ - \dfrac{a^{4}}{45\pi ^{2}}
\left[  \lambda \widetilde{f}_{1}(a\beta ) 
    +15 f_{3}(a\beta ) + 10\lambda\widetilde{f}_{3}(a\beta )
    \right. \nonumber\\ & \qquad \left . + 4\lambda \widetilde{f}_{5}(a\beta )\right].  
    \label{eq:Tcoeffs}
\end{align}
\end{subequations}

Since the components of the RSET have Taylor series expansions (\ref{eq:RSETseries}) for small $\rho $ which contain only even powers of $\rho $, we assume that the same  is true of the QC metric functions $A(\rho )$, $B(\rho )$ and $C(\rho )$: 
\begin{subequations}
\label{eq:ABCseries}
\begin{align}
    A(\rho ) & = A_{0} + A_{2}\rho ^{2}+ {\mathcal {O}}(\rho ^{4}) ,
    \label{eq:Aseries}
    \\
    B(\rho ) & = B_{0} + B_{2}\rho ^{2} +{\mathcal {O}} (\rho ^{4}) ,
    \label{eq:Bseries} 
    \\
    C(\rho ) & = C_{2}\rho ^{2} + C_{4} \rho ^{4} + {\mathcal {O}} (\rho ^{6}) ,
    \label{eq:Cseries}
\end{align}
\end{subequations}
where $A_{i}$, $B_{i}$, $C_{i}$ are constant coefficients. 
In (\ref{eq:ABCseries}), we have assumed that the QC metric function $C(\rho )$ is ${\mathcal {O}}(\rho ^{2})$ as $\rho \rightarrow 0$, which is the same behaviour as in the original AdS metric (\ref{eq:pureAdSmetric}). 

We additionally assume that the metric (\ref{eq:QCadSansatz}) has a regular origin at $\rho = 0$. 
In order to avoid a curvature singularity at $\rho =0$, it must be the case that
\begin{equation}
    B_{0}=C_{2}.  
    \label{eq:B2C0}
\end{equation}
Since we have assumed that the QC metric (\ref{eq:QCadSansatz}) is static, we can rescale the time coordinate $\tau $, which leads to a rescaling of the metric function $A(\rho )$.
For this reason the coefficient $A_{0}$ in (\ref{eq:Aseries}) is not determined by the QCEE  (\ref{eq:Aeqn}). 
We can therefore set
\begin{equation}
    A_{0}=1
    \label{eq:A0}
\end{equation}
without loss of generality.
All the remaining constants in (\ref{eq:ABCseries}) can be fixed by substituting the series (\ref{eq:RSETseries}, \ref{eq:ABCseries}) into the QCEE (\ref{eq:Aeqn}, \ref{eq:Beqn}, \ref{eq:singleCeqn}) and ensuring that these equations are satisfied order-by-order in $\rho $.
The lowest order term in (\ref{eq:Aeqn}) gives
\begin{equation}
    A_{2} = \dfrac{2\left( 2 T_{\rho 2}- T_{\theta 2} \right)}{ T_{\tau 0 } -T_{\theta 0}},
\end{equation}
which is finite since, from (\ref{eq:RSETseries}), we have $T_{\tau 0 } \neq T_{\theta 0}$.
From the lowest-order term in (\ref{eq:singleCeqn}), we find
\begin{equation}
    C_{2} = \dfrac{12\left( 2 T_{\theta 2} - T_{\rho 2} \right) }{\left(  T_{\tau 0 } -T_{\theta 0} \right) \left(  T_{\tau 0 } -3T_{\theta 0 } + 2{\widetilde {\Lambda }} \right) },
    \label{eq:C2}
\end{equation}
where both terms in the denominator are nonvanishing.

Following the above procedure, we can find series expansions of the RSET and QC metric functions to any desired order in $\rho $, although the algebraic expressions for the coefficients in the QC metric functions rapidly become extremely complicated.

Our numerical method begins by integrating the nonlinear equation for $C(\rho )$ (\ref{eq:singleCeqnFull}), using {\tt {Mathematica}}'s inbuilt {\tt {NDSolveValue}} command.
We begin the numerical integration at $\rho = \epsilon = 10^{-3}$. 
Initial conditions for the numerical integration are given by the series expansion (\ref{eq:Cseries}).  
We perform a series expansion in the RSET components up to  ${\mathcal {O}}(\rho ^{8})$, which is sufficient to give series in $A(\rho )$ and $B(\rho )$ up to ${\mathcal {O}}(\rho ^{6})$ and in $C(\rho )$ up to ${\mathcal {O}}(\rho ^{8})$. 
The coefficients in the series expansions of the RSET components (\ref{eq:RSETseries}) involve infinite sums which converge extremely rapidly; we included just the first 100 terms in each sum, which was sufficient for our purposes. 
With series expansions up to the orders given above, and taking $\epsilon $ in the range $10^{-2}-10^{-3}$ we estimate the relative error in the solutions of the QCEE to be approximately $10^{-8}$.

The {\tt {NDSolveValue}} command gives $C(\rho )$ as an interpolating function. 
From this, the QC metric function $B(\rho )$ is given directly by (\ref{eq:Beqn}), without any further integration.
To find the remaining QC metric function $A(\rho )$, we again apply the {\tt {NDSolveValue}} command to numerically integrate (\ref{eq:Aeqn}) from $\rho = \epsilon $, using the series (\ref{eq:Aseries}) up to ${\mathcal {O}}(\rho ^{6})$ as the initial condition.
In all our numerical integration, we end the integration at $\rho = \tfrac{\pi }{2} - \epsilon $ since the QC metric functions $A(\rho )$, $B(\rho )$ and $C(\rho )$ all diverge as $\rho \rightarrow \tfrac{\pi }{2}$ (as is the case for the original AdS metric functions (\ref{eq:pureAdSmetric})).
In all our numerical computations we have set the original inverse AdS  length scale $a$ to be equal to unity; this means the the renormalized inverse AdS length scale is
\begin{equation}
    {\widetilde {a}} = {\sqrt { -\dfrac{{\widetilde {\Lambda }}}{3}}} =  0.9999824 \quad {\mbox {to 7 significant figures}},
\end{equation}
where the renormalized cosmological constant ${\widetilde {\Lambda }}$ is given by (\ref{eq:adSLren}).

\section{QC AdS metrics}
\label{sec:ABC}

In this section we present our results for the QC AdS metrics found by solving the QC Einstein equations, as discussed in the previous section. 
We first consider the solutions when Dirichlet or Neumann boundary conditions have been applied to the quantum scalar field, before turning to the case of transparent boundary conditions.
Our main interest in this section is to explore the deviation of the QC metrics from pure AdS space-time, as well as 
considering the effect the boundary conditions have on the QC metrics.

\subsection{Dirichlet and Neumann boundary conditions}
\label{sec:DNBCs}

\begin{figure}
  \includegraphics[width=\linewidth]{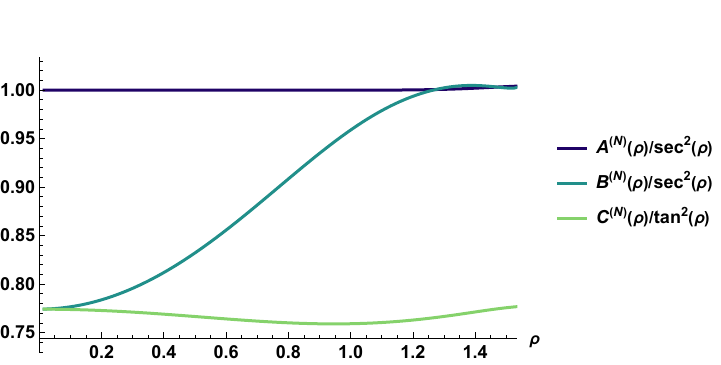}
\caption{Ratios between the components of the QC metric (\ref{eq:QCadSansatz}) and the pure AdS metric (\ref{eq:pureAdSmetric}) for Neumann boundary conditions. We have fixed the (unrenormalized) inverse AdS length scale $a=1$ and the inverse temperature $\beta=\tfrac{\pi}{4}$.}
  \label{fig:NratioPlotABCtemphalf}
\end{figure}

In Fig.~\ref{fig:NratioPlotABCtemphalf} we show a typical QC metric solution. We have fixed the inverse temperature in the RSET to be $\beta = \tfrac{\pi }{4}$ and applied Neumann boundary conditions to the quantum field.
We see that the metric function $A^{(\rm {N})}(\rho )$ is indistinguishable from that in pure AdS when plotted on this scale, while the metric functions $B^{(\rm {N})}(\rho )$ and $C^{(\rm {N})}(\rho )$ are significantly different from the corresponding functions in pure AdS. 
At the origin $\rho =0$, we have fixed $A^{(\rm {N})}(\rho ) =1$ as seen in Fig.~\ref{fig:NratioPlotABCtemphalf}, and we also see that $B^{(\rm {N})}(\rho )/\sec ^{2} \rho $ and $C^{(\rm {N})}(\rho )/\tan ^{2} \rho $ have the same value, in accordance with the series expansions (\ref{eq:Bseries}, \ref{eq:Cseries}, \ref{eq:B2C0}). 
The function $B^{(\rm {N})}(\rho )/\sec ^{2} \rho $ is monotonically increasing from the origin towards the boundary at $\rho = \pi /2$, except for a small region close to the boundary where it is monotonically decreasing. 
In contrast, the function $C^{(\rm {N})}(\rho )/\tan ^{2} \rho $ is decreasing close to the origin, has a local minimum and then increases close to the boundary.
Typical solutions for Dirichlet boundary conditions show similar behaviour.

\begin{figure}
\centering
\begin{subfigure}{\columnwidth}
  \centering
  \includegraphics[width=\columnwidth]{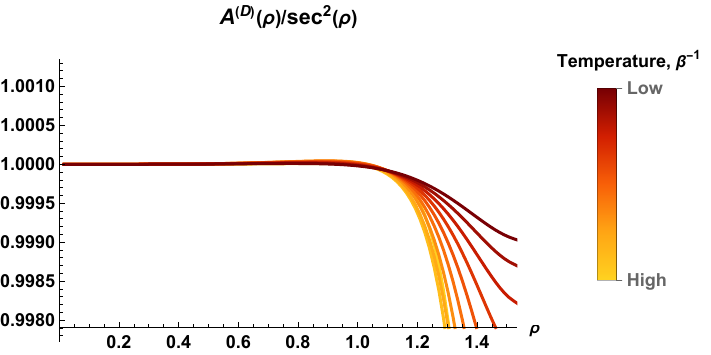}
  \label{fig:DratioPlotAtempvaried}
\end{subfigure}%
\\[0.3cm]
\begin{subfigure}{\columnwidth}
  \centering
  \includegraphics[width=\columnwidth]{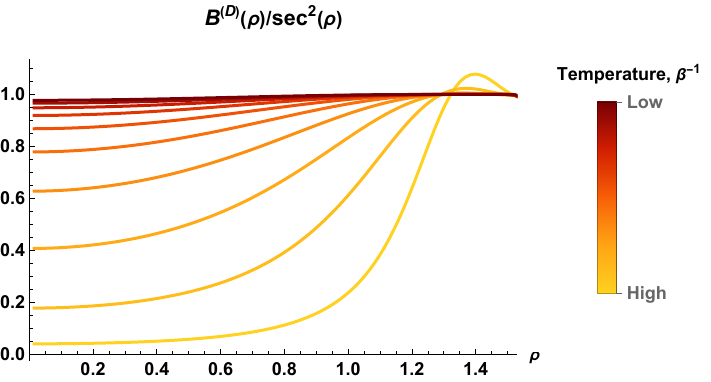}
  \label{fig:DratioPlotBtempvaried}
\end{subfigure}
\\[0.3cm]
\begin{subfigure}{\columnwidth}
  \centering
  \includegraphics[width=\columnwidth]{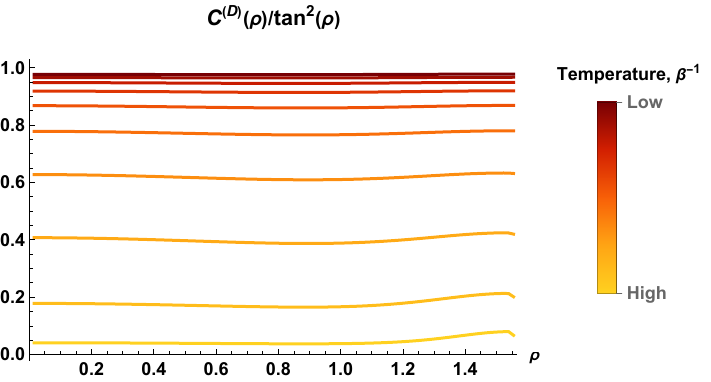}
  \label{fig:DratioPlotCtempvaried}
\end{subfigure}
\caption{Ratios between the components of the QC metric (\ref{eq:QCadSansatz}) and the pure AdS metric (\ref{eq:pureAdSmetric}) for Dirichlet boundary conditions.
The top plot shows $A^{(\rm {D})}(\rho )/\sec ^{2}\rho $, the middle plot $B^{(\rm {D})}(\rho )/\sec ^{2}\rho $ and the bottom plot $C^{(\rm {D})}(\rho )/\tan ^{2} \rho $. Keeping the inverse AdS length scale $a=1$ fixed, the inverse temperature lies in the range $\beta \in (\frac{\pi}{12},\frac{11\pi}{24})$.}
\label{fig:DNratioPlotsABCtempvaried}
\end{figure}

We now study how the QC metrics change as we vary the inverse temperature $\beta $ in the RSET. 
In Fig.~\ref{fig:DNratioPlotsABCtempvaried} we plot the ratios $A^{(\rm {D})}(\rho )/\sec ^{2}\rho $ (top plot), $B^{(\rm {D})}(\rho )/\sec ^{2} \rho $ (middle plot) and $C^{(\rm {D})}(\rho )/\tan ^{2}(\rho ) $ (bottom plot)  for Dirichlet boundary conditions. 
We have found QC metrics for a range of inverse temperatures, with $\beta = \tfrac{\pi }{12}$ corresponding to the highest temperature studied. 
The zero-temperature limit corresponds to $\beta \rightarrow \infty $; in this case the QC metric is just pure AdS with a renormalized cosmological constant (\ref{eq:adSLren}). 

First consider the ratio $A^{(\rm {D})}(\rho )/\sec^{2} \rho $ (top plot in Fig.~{\ref{fig:DNratioPlotsABCtempvaried}}). 
Looking at the vertical scale, it can be seen that this ratio is very close to unity for all values of the temperature considered and all $\rho \in [0,\tfrac{\pi }{2})$. 
The only significant deviation from unity is close to the boundary as $\rho \rightarrow \tfrac{\pi }{2}$, 
with the deviations from unity increasing as the temperature increases.

Turning now to the ratio $B^{(\rm {D})}(\rho )/\sec ^{2} \rho $ (middle plot in Fig.~\ref{fig:DNratioPlotsABCtempvaried}), we find much more significant variations compared to the pure AdS case. 
At the origin, the value of the ratio $B^{(\rm {D})}(\rho )/\sec ^{2} \rho $ decreases rapidly as the temperature $\beta ^{-1}$ increases, and is close to zero for the large temperature $\beta ^{-1} = \tfrac{12}{\pi }$. 
For all values of the inverse temperature studied, we find that the ratio $B^{(\rm {D})}(\rho )/\sec ^{2}\rho  $ initially increases as $\rho $ increases from zero.
The ratio $B^{(\rm {D})}(\rho )/\sec ^{2} \rho $ then has a maximum for $\rho $ close to the boundary.
For sufficiently high temperature, the maximum value of the ratio $B^{(\rm {D})}(\rho )/\sec ^{2} \rho $ is greater than unity.
To within the accuracy of our numerical integration, it appears to be the case that $B^{(\rm {D})}(\rho )/\sec ^{2}\rho $ approaches unity as $\rho \rightarrow \tfrac{\pi}{2}$ and the space-time boundary is approached. 

Finally we consider the ratio $C^{(\rm {D})}(\rho )/\tan ^{2} \rho $ (bottom plot in Fig.~\ref{fig:DNratioPlotsABCtempvaried}).
Like the ratio $B^{(\rm {D})}(\rho )/\sec ^{2} \rho $,  at the origin $\rho =0$ the ratio $C^{(\rm {D})}(\rho )/\tan ^{2}\rho $ decreases rapidly as the temperature increases, and is close to zero for the highest temperature studied (corresponding to $\beta = \tfrac{\pi }{12}$). 
Unlike the ratio $B^{(\rm {D})}(\rho )/\sec ^{2} \rho $, we see that $C^{(\rm {D})}(\rho )/\tan ^{2} \rho $ does not change greatly as the radial coordinate $\rho $ increases.  The ratio generally has a local minimum for some value of $\rho $ and is slightly increasing close to the space-time boundary.
The value that the ratio $C^{(\rm {D})}(\rho )/\tan ^{2} \rho $ approaches as $\rho \rightarrow \tfrac{\pi }{2}$ decreases with increasing temperature. 

\begin{figure}
\centering
\begin{subfigure}{\columnwidth}
  \centering
  \includegraphics[width=\columnwidth]{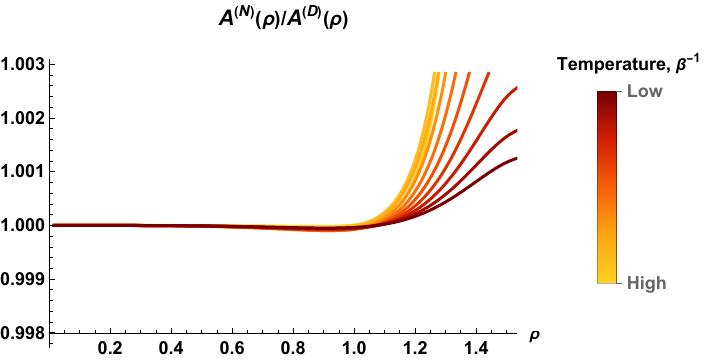}
  \label{fig:NratioPlotAtempvaried}
\end{subfigure}%
\\
\begin{subfigure}{\columnwidth}
  \centering
  \includegraphics[width=\columnwidth]{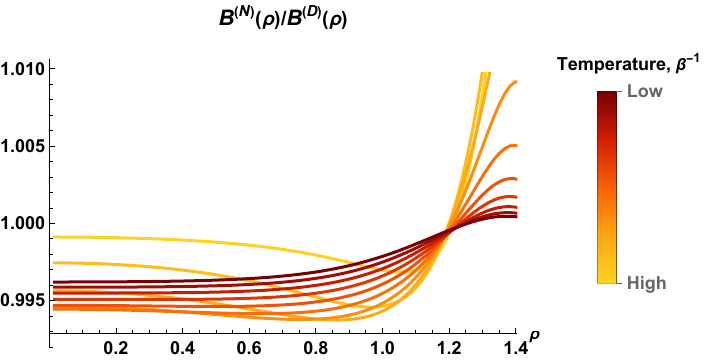}
  \label{fig:NratioPlotBtempvaried}
\end{subfigure}
\\
\begin{subfigure}{\columnwidth}
  \centering
  \includegraphics[width=\columnwidth]{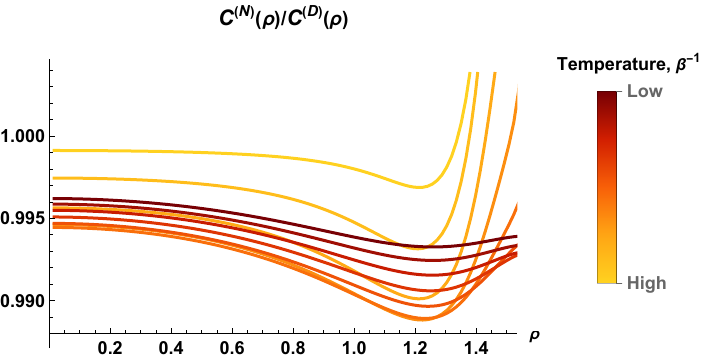}
  \label{fig:NratioPlotCtempvaried}
\end{subfigure}
\caption{Ratios between the components of the QC metric (\ref{eq:QCadSansatz}) when Neumann boundary conditions are applied to the scalar field and the corresponding QC metric components when Dirichlet boundary conditions are applied.
The top plot shows $A^{(\rm N)}(\rho )/A^{(\rm D)}(\rho )$, the middle plot $B^{(\rm N)}(\rho )/B^{(\rm D)}(\rho )$ and the bottom plot $C^{(\rm N)}(\rho )/C^{(\rm D)}(\rho )$. Keeping the inverse AdS length scale $a=1$ fixed, the inverse temperature lies in the range $\beta \in (\frac{\pi}{12},\frac{11\pi}{24})$.}
\label{fig:NratioPlotsABCtempvaried}
\end{figure}

When Neumann boundary conditions are applied to the scalar field, the plots of the ratios of the QC metric components to the AdS metric components $B^{(\rm N)}(\rho )/\sec ^{2}\rho $ and $C^{(\rm N)}(\rho )/\tan ^{2}(\rho )$ are virtually indistinguishable by eye from the corresponding plots for Dirichlet boundary conditions, shown in Fig.~\ref{fig:DNratioPlotsABCtempvaried}.
In Fig.~\ref{fig:NratioPlotsABCtempvaried} we therefore show the ratios of the QC metric components when Neumann boundary conditions are applied to the corresponding QC metric components when Dirichlet boundary conditions are applied. 
From these ratios, we see that the difference between the QC metric functions for the different boundary conditions is very small, typically less than one percent. 

When we examine the data for the ratio $A^{(\rm {N})}(\rho )/\sec ^{2}(\rho )$, while this is less than unity close to the boundary when Dirichlet boundary conditions are applied, it is greater than unity for Neumann boundary conditions. Furthermore, the deviations from unity are larger for Neumann boundary conditions compared with Dirichlet boundary conditions, and become significant at smaller values of $\rho $ for Neumann boundary conditions compared with Dirichlet.

From the plots of the ratios of the QC metric functions $B(\rho )$ and $C(\rho )$ shown in Fig.~\ref{fig:NratioPlotsABCtempvaried}, close to the origin at $\rho =0$ these ratios are closer to unity for higher temperatures. 
Thus the effect of the boundary conditions on the QC metric is less significant at higher temperatures.
This is in accordance with the properties of the RSET \cite{Morley:2023exv}, whose components at high temperature show little dependence on the boundary conditions applied to the scalar field.
Close to the space-time boundary, the QC metric ratios in Fig.~\ref{fig:NratioPlotsABCtempvaried} are significantly greater than unity, and increase with increasing temperature, mimicking the behaviour of the RSET close to the boundary in Fig.~\ref{fig:EnergyRatio}.

\subsection{Transparent boundary conditions}
\label{sec:transparent}

When transparent boundary conditions are applied to the quantum scalar field, the RSET (\ref{eq:RSET}) simplifies considerably: 
\begin{equation}
    {\widetilde {T}}_{\mu }{}^{\nu }
    = \dfrac{a^{2}}{6\pi ^{2}} \cos ^{4} \rho \, f_{3}(a\beta ) \, {\rm {Diag}}  \, \{
    -3, 1, 1, 1 
    \} ,
    \label{eq:RSETtrans}
\end{equation}
where $f_{3}(a\beta )$ is given in (\ref{eq:InfSumfAFG}).
In this case the equation for $A(\rho )$ (\ref{eq:Aeqn}) reduces to
\begin{equation}
    \dfrac{A'}{A} =2 \tan \rho ,
\end{equation}
which can be directly integrated to give
\begin{equation}
    A(\rho )  = \sec ^{2} \rho ,
    \label{eq:Atrans}
\end{equation}
where we have chosen a constant of integration such that $A(0)=1$, as in (\ref{eq:A0}). 
Therefore the QC metric function $A^{(\rm {T})}(\rho )$ takes exactly the same form as in pure AdS.

The QC metric functions $B^{(\rm {T})}(\rho )$ and $C^{(\rm {T})}(\rho )$ are, however, not the same as for pure AdS.
The equations governing these (\ref{eq:Beqn}, \ref{eq:singleCeqn}) do however simplify, giving the following equation for $C(\rho )$:
\begin{subequations}
\label{eq:Ceqntransfull}
\begin{align} 
0 & = 
\frac{\Xi_\tau }{C}
  + \dfrac{4}{CC'}\left( 2CC'' + C'^2 \right) \dfrac{\Xi_\rho {\widetilde {T}}_{\rho}{}^{\rho}}{\Upsilon} 
+8\dfrac{\diff}{\diff{\rho}}\left(\dfrac{\Xi_\rho  {\widetilde {T}}_{\rho}{}^{\rho} }{\Upsilon}\right), 
   \label{eq:Ceqntrans}  
\end{align}
where $\Upsilon $  (\ref{eq:Upsilon}) is now
\begin{equation}
    \Upsilon =
    4  \left( C
   {\widetilde {T}}_{\rho}{}^{\rho} {}'
    -C' {\widetilde {T}}_{\rho }{}^{\rho } \right) ,
\end{equation}
and $\Xi_\bullet $ (\ref{eq:Xi}) is unchanged.
\end{subequations}
The equation for $B(\rho )$ in terms of $C(\rho )$ (\ref{eq:Beqn}) also simplifies considerably:
\begin{equation}
    B(\rho ) = 
    -\dfrac{C'}{4C\Xi_\rho } \left( C'+ 4C \tan \rho \right) .
    \label{eq:Btrans}
\end{equation}

\begin{figure}
\centering
\begin{subfigure}{\columnwidth}
  \centering
  \includegraphics[width=\columnwidth]{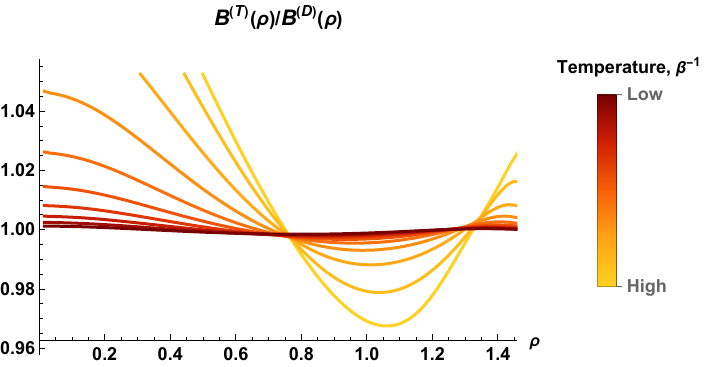}
  \label{fig:TratioPlotBtempvaried}
\end{subfigure}
\\
\begin{subfigure}{\columnwidth}
  \centering
  \includegraphics[width=\columnwidth]{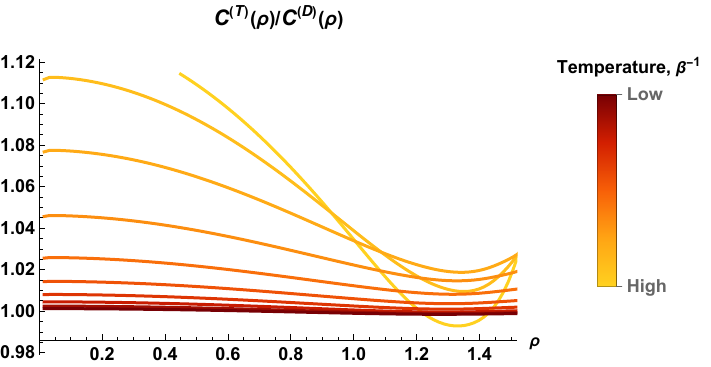}
  \label{fig:TratioPlotCtempvaried}
\end{subfigure}
\caption{Ratios between the components of the QC metric (\ref{eq:QCadSansatz}) when transparent boundary conditions are applied to the scalar field and the corresponding QC metric components when Dirichlet boundary conditions are applied.
The upper plot shows $B^{(\rm T)}(\rho )/B^{(\rm D)}(\rho )$ and the lower plot $C^{(\rm T)}(\rho )/C^{(\rm D)}(\rho )$. Keeping the inverse AdS length scale $a=1$ fixed, the inverse temperature lies in the range $\beta \in (\frac{\pi}{12},\frac{11\pi}{24})$.}
\label{fig:TratioPlotsABCtempvaried}
\end{figure}

Although the equations governing $C(\rho)$ (\ref{eq:Ceqntransfull}) and $B(\rho )$ (\ref{eq:Btrans}) are much simpler than for either Dirchlet or Neumann boundary conditions, the resulting metric functions $B^{(\rm {T})}(\rho )$ and $C^{(\rm {T})}(\rho )$ are very similar to those shown in Fig.~\ref{fig:DNratioPlotsABCtempvaried}. 
In particular, we find that $B^{(\rm {T})}(\rho )$ and $C^{(\rm {T})}(\rho )$ for transparent boundary conditions are not the same as for pure AdS, even though the metric function $A^{(\rm {T})}(\rho )$ (\ref{eq:Atrans}) does take the AdS form.
Fig.~\ref{fig:TratioPlotsABCtempvaried} shows the ratios of the QC metric functions when transparent boundary conditions are applied to those for Dirichlet boundary conditions, namely $B^{(\rm T)}(\rho)/B^{(\rm D)}(\rho)$ (upper plot) and $C^{(\rm T)}(\rho)/C^{(\rm D)}(\rho)$ (lower plot). 
At low temperatures, the QC metric functions $B^{(\rm T)}(\rho)$ and $C^{(\rm T)}(\rho )$ for transparent boundary conditions are indistinguistable from the corresponding QC metric functions for Dirichlet boundary conditions.
At higher temperatures, the ratios $B^{(\rm T)}(\rho)/B^{(\rm D)}(\rho)$  and $C^{(\rm T)}(\rho)/C^{(\rm D)}(\rho)$ are significantly different from unity (more so than the ratios  $B^{(\rm N)}(\rho)/B^{(\rm D)}(\rho)$  and $C^{(\rm N)}(\rho)/C^{(\rm D)}(\rho)$ shown in Fig.~\ref{fig:NratioPlotsABCtempvaried}). 

To derive the vacuum (and hence thermal) Green's function with transparent boundary conditions applied to the conformally-coupled scalar field on AdS, we first make a conformal transformation to
the Einstein static universe (ESU) with metric
\begin{equation}
    \diff{{\widetilde {s}}}^{2} = a^{-2}\left[ -\diff{\tau}^2 + \diff{\rho}^2 +\sin^2\rho\diff{\Omega}^2 \right] .
    \label{eq:ESU}
\end{equation}
Since ESU is a globally-hyperbolic space-time, it is not necessary to apply boundary conditions to the field on ESU. 
Having found the vacuum Green's function on ESU, the corresponding Green's function on AdS will satisfy transparent boundary conditions \cite{Avis:1977yn}.
The resulting vacuum Green's function on AdS is then used to construct the corresponding thermal Green's function on AdS, from which the RSET (\ref{eq:RSETtrans}) is then obtained \cite{Allen:1986ty}. 
It is therefore informative to compare our results in this section for the QC AdS metric with the corresponding QC ESU metric.
ESU solutions of the SCEE (\ref{eq:semiclassicalEinEqns}) are studied in detail in \cite{Sanders:2020osl}.
Our analysis below is rather simpler, as we are considering a massless, conformally-coupled scalar field and are only interested in the QC metric rather than solving the full SCEE.

Unlike AdS, the ESU is not a maximally-symmetric space-time, and so the general approach in Sec.~\ref{sec:maxsym} is not immediately applicable. 
The ESU is a solution of the classical Einstein equations (\ref{eq:classicalEinEqns}) with a cosmological constant $\Lambda = a^{2}$ and a nonzero classical stress-energy tensor 
\begin{equation}
    {}^{\rm {dust}}T_{\mu }{}^{\nu } = -2a^{2} \, {\rm {Diag}} \{ 1, 0, 0, 0 \} ,
    \label{eq:dust}
\end{equation}
corresponding to pressureless dust. 
Following \cite{Sanders:2020osl}, for the rest of this section only we reinstate Newton's constant and consider the modified QCEE
\begin{equation}
    G_{\mu}{}^{\nu} + \Lambda \delta _{\mu}{}^{\nu} = 8\pi G \left[ {}^{\rm {dust}}T_{\mu }{}^{\nu } + \langle {\hat {T}}_{\mu}{}^{\nu} \rangle  ^{\rm {ESU}} \right]
    \label{eq:ESUSCEE}
\end{equation}
where $\langle {\hat {T}}_{\mu}{}^{\nu} \rangle  ^{\rm {ESU}}$ is the RSET for the quantum scalar field on ESU and we have raised the second index for later convenience.
In (\ref{eq:ESUSCEE}),  we have also set the constants $\alpha  $ and $\beta $ in (\ref{eq:semiclassicalEinEqns}) equal to zero; this will be justified {\it {a posteriori}} later in this section.

The RSET for a massless, conformally coupled scalar field on ESU  $\langle {\hat {T}}_{\mu }{}^{\nu } \rangle ^{\rm {ESU}}$ is related to that on AdS $\langle {\hat {T}}_{\mu }{}^{\nu } \rangle ^{\rm {AdS}}$ in the corresponding state by \cite{Birrell:1982ix}
\begin{equation}
    \langle {\hat {T}}_{\mu }{}^{\nu } \rangle ^{\rm {AdS}} = 
    \langle {\hat {T}}_{\mu }{}^{\nu } \rangle ^{\rm {ESU}} \dfrac{{\sqrt {-{\widetilde {g}}}}}{{\sqrt {-g}}} 
    - \dfrac{1}{2880 \pi ^{2}}H_{\mu }^{(3)\nu } ,
    \label{eq:confrel}
\end{equation}
where $g$, ${\widetilde {g}}$ are, respectively, the determinants of the metrics (\ref{eq:pureAdSmetric}, \ref{eq:ESU}), and
\begin{align}
    H_{\mu \nu }^{(3)}
     = & \frac{1}{3}\nabla _{\mu }\nabla _{\nu }R-\frac{1}{3}g_{\mu \nu }\nabla _{\sigma }\nabla ^{\sigma }R - \frac{1}{6}g_{\mu \nu }R^{2}+\frac{1}{3}RR_{\mu \nu }
     \nonumber \\ & \qquad
    +R^{\rho \sigma }R_{\rho \mu \sigma \nu } .
\end{align}
Since the second term in (\ref{eq:confrel}) is independent of the quantum state, we have
\begin{equation}
{}^{\rm {AdS}}{\widetilde {T}}_{\mu }{}^{\nu }
    =  {}^{\rm {ESU}}{\widetilde {T}}_{\mu }{}^{\nu }\cos ^{4} \rho ,
\end{equation}
where ${}^{\rm {AdS}}{\widetilde {T}}_{\mu }{}^{\nu }$ is the difference between the thermal and vacuum expectation values on AdS with transparent boundary conditions (\ref{eq:RSETtrans}) and 
\begin{equation}
    {}^{\rm {ESU}}{\widetilde {T}}_{\mu }{}^{\nu } =    \langle \beta \lvert  {\hat {T}}_{\mu}{}^{\nu} \rvert \beta  \rangle ^{\rm {ESU}}
    -  \langle 0 \lvert  {\hat {T}}_{\mu}{}^{\nu} \rvert 0  \rangle ^{\rm {ESU}} .
\end{equation}
Therefore the difference between the thermal and vacuum expectation values of the RSET on ESU takes a particularly simple form:
\begin{equation}
    {}^{\rm {ESU}}{\widetilde {T}}_{\mu }{}^{\nu }
    = \dfrac{a^{2}}{6\pi ^{2}} \, f_{3}(a\beta ) \, {\rm {Diag}}  \, \{
    -3, 1, 1, 1 
    \} .
    \label{eq:ESUthermal}
\end{equation}
It is striking that (\ref{eq:ESUthermal}) is proportional to the RSET in the vacuum state on ESU 
\cite{Candelas:1978gf}:
\begin{equation}
     \langle 0_{{\rm {ESU}}} | {\hat {T}}_{\mu}{}^{\nu} | 0 _{{\rm {ESU}}} \rangle 
     = 
    \dfrac{a^{2}}{480\pi ^{2}} {\rm {Diag}} \{ -3, 1,1,1 \} .
    \label{eq:RSETESU}
\end{equation}
Hence, to find the QC ESU metric, we  consider the RSET
\begin{equation}
     \langle {\hat {T}}_{\mu }{}^{\nu } \rangle ^{\rm {ESU}}
    = a^{2}{\mathcal {P}} \, {\rm {Diag}}  \, \{
    -3, 1, 1, 1 
    \} ,
    \label{eq:ESUgeneral}
\end{equation}
where the constant ${\mathcal {P}}$ is given by
\begin{equation}
    {\mathcal {P}} = \frac{1}{480\pi ^{2}} \left[ 1+80f_{3}(a\beta ) \right] .
\end{equation}

To solve (\ref{eq:ESUSCEE}) with (\ref{eq:ESUgeneral}) on the right-hand-side, we first define a renormalized cosmological constant ${\widetilde {\Lambda }}$  (and hence a renormalized inverse length scale ${\widetilde {a}}$) using the spatial parts of the total RSET:
\begin{equation}
    {\widetilde {\Lambda }} = {\widetilde {a}}^{2} = \Lambda - 8\pi G a^{2}{\mathcal {P}} = a^{2} \left( 1 - 8\pi G {\mathcal {P}} \right). 
\end{equation}
The modified SCEE (\ref{eq:ESUSCEE}) now read
\begin{equation}
    G_{\mu }{}^{\nu } + {\widetilde {\Lambda }}\delta _{\mu }{}^{\nu } = -16\pi G a^{2}  \, {\rm {Diag}}  \, \{
    1+2{\mathcal {P}}, 0, 0, 0 
    \}  .
    \label{eq:ESUSCEE1}
\end{equation}
The right-hand-side of (\ref{eq:ESUSCEE1}) now has the pressure dust form (\ref{eq:dust}) with renormalized inverse length scale ${\widetilde {a}}$ if we also define a renormalized Newton constant ${\widetilde{G}}$ (as in \cite{Sanders:2020osl}) by 
\begin{equation}
    8\pi {\widetilde {G}} = 8\pi G \dfrac{a^{2}}{{\widetilde {a}}^{2}} \left( 1 + 2{\mathcal {P}} \right) 
    = \dfrac{8\pi G(1+2{\mathcal {P}})}{1-8\pi G{\mathcal {P}}}
    = \dfrac{1+2{\mathcal {P}}}{1-{\mathcal {P}}},
\end{equation}
where in the last equality we have set $8\pi G=1$, as elsewhere in this paper.
We therefore obtain the QCEE
\begin{equation}
    G_{\mu \nu }^{\rm {QC}}+ {\widetilde {\Lambda }} g_{\mu \nu }^{\rm {QC}} = - 2{\widetilde {a}}^{2}  {\rm {Diag}} \{ 1, 0, 0, 0 \}, 
\end{equation}
and hence the QC metric takes the ESU form (\ref{eq:ESU}), with renormalized inverse length scale ${\widetilde {a}}$. 

Therefore, providing we renormalize both the cosmological constant ${\widetilde {\Lambda }}$ and Newton's constant ${\widetilde {G}}$, we are justified in setting the constants $\alpha  $ and $\beta $ equal to zero in (\ref{eq:ESUSCEE}).
While the AdS and ESU space-times are conformally related, and we are dealing with a conformal quantum field, it is not the case that the resulting QC metrics are also conformally related, due to the nonlinearity of the QCEE. 
If the scalar field is in a thermal state on ESU, the resulting QC metric has ESU form, but the QC metric sourced by the corresponding thermal state on AdS is not of the AdS form.

\section{QC AdS solitons}
\label{sec:solitons}

In the previous section we have constructed the QC AdS metrics by integrating the QCEE (\ref{eq:QCEE}).
We have found that, in the low-temperature limit, the QC AdS metrics approach pure AdS space-time, with deviations from AdS increasing as the temperature of the thermal quantum state increases.
In this section we consider an alternative metric ansatz for the QC AdS metrics, which will aid their interpretation as solitons sourced by the quantum scalar field.

\subsection{Alternative QCEE form}
\label{eq:altQCEE}

To transform the QC AdS metric ansatz (\ref{eq:QCadSansatz}) into the more familiar form for a static, spherically symmetric space-time, we first make a change of radial coordinate, defining a new coordinate $r$ by
\begin{equation}
    r^{2} = C(\rho ).
    \label{eq:rdef}
\end{equation}
From the results of the previous section, we see that $r\rightarrow 0$ as $\rho \rightarrow 0 $ from (\ref{eq:Cseries}). 
Since $C(\rho )/\tan ^{2}\rho $ tends to a finite nonzero limit as $\rho \rightarrow \pi /2$ (see Fig.~\ref{fig:DNratioPlotsABCtempvaried}), we also have $r\rightarrow \infty $ as $\rho \rightarrow \pi /2$, so that the coordinate $r$ has range $r\in [0, \infty )$ as is usually the case for the radial coordinate in spherical polars.

We then make the following metric ansatz:
\begin{equation}
    \diff{s}^2 = -f(r)e^{\psi(r)}\diff{t}^2 + f(r)^{-1}\diff{r}^2 + r^2\diff{\Omega}^2,
    \label{eq:sphmetric}
\end{equation}
in terms of functions $f(r)$ and $\psi (r)$ which are to be determined from the QCEE. 
We also write $f(r)$ as follows:
\begin{equation}
\label{eq:fDefSCEErForm}
   f(r) = 1 - \dfrac{2m(r)}{r} - \dfrac{{\widetilde {\Lambda}}}{3}r^2 ,
\end{equation}
where ${\widetilde {\Lambda }}$ is the renormalized cosmological constant (\ref{eq:adSLren}) and $m(r)$ is to be determined from the QCEE.
Pure AdS space-time has $m(r) \equiv 0 $.
Differentiating (\ref{eq:rdef}) and comparing the two forms (\ref{eq:QCadSansatz}, \ref{eq:sphmetric}) of the QC AdS metric, we have 
\begin{align}
    f(r) & = \dfrac{C'(\rho)^2}{4B(\rho)C(\rho)} ,
\notag 
\\
    m(r) &= \dfrac{\sqrt{C(\rho)}}{2}\left(1 + {\widetilde {a}}^2C(\rho) - \dfrac{C'(\rho)^2}{4B(\rho)C(\rho)}\right), 
    \notag 
    \\ e^{\psi (r)} &= \dfrac{4A(\rho)B(\rho)C(\rho)}{C'(\rho)^2} ,
    \label{eq:fmpsi}
\end{align}
where ${\widetilde {a}}$ is the renormalized inverse AdS length scale. 
If we attempt to find $f(r)$, $m(r)$ and $\psi (r)$ using the expressions (\ref{eq:fmpsi}), we find that numerical errors in the metric functions $A(\rho )$, $B(\rho )$ and $C(\rho )$ accumulate rapidly.
To obtain numerically satisfactory results for $f(r)$, $m(r)$ and $\psi (r)$ we instead integrate the QCEE resulting from the metric ansatz (\ref{eq:sphmetric}).

The QCEE for the metric functions $m(r)$ and $\psi (r)$ take a particularly simple form:
\begin{subequations}
\label{eq:QCEEsphsym}
    \begin{align}
    \dfrac{\diff{m}}{\diff {r}} & = ~   - \dfrac{r^{2}}{2}  {\widetilde {T}}_{\tau}{}^{\tau},
    \label{eq:meqn}
    \\
    \dfrac{\diff{\psi }}{\diff {r}}  & = ~  -\dfrac{r}{f(r)} \left( 
    {\widetilde {T}}_{\tau}{}^{\tau}
    -  {\widetilde {T}}_{\rho}{}^{\rho} 
    \right) .
    \label{eq:psieqn}
\end{align}
\end{subequations}
Our primary interest lies in the metric function $m(r)$, which we consider in the following subsection.

\subsection{QC AdS soliton masses}
\label{sec:mass}

\begin{figure}
    \centering
\begin{subfigure}{\columnwidth}
    \includegraphics[width=\columnwidth]{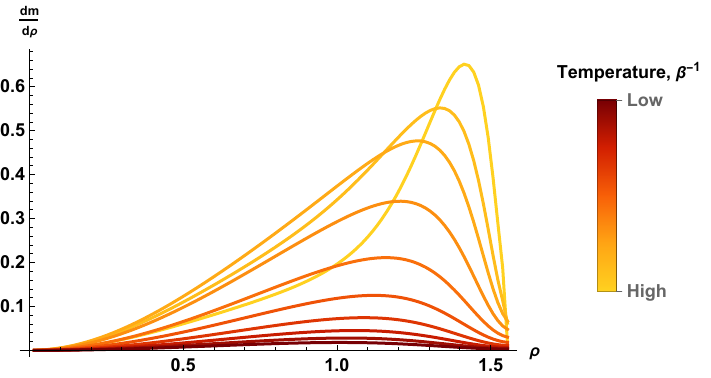}
    \caption{Dirichlet boundary conditions}
    \label{fig:dmdrhoDirichlet}
\end{subfigure}
\begin{subfigure}{\columnwidth}
    \includegraphics[width=\columnwidth]{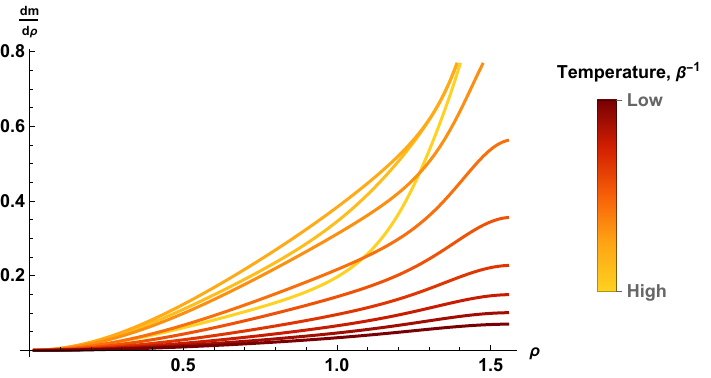}
    \caption{Neumann boundary conditions}
    \label{fig:dmdrhoNeumann}
\end{subfigure}
\begin{subfigure}{\columnwidth}
    \includegraphics[width=\columnwidth]{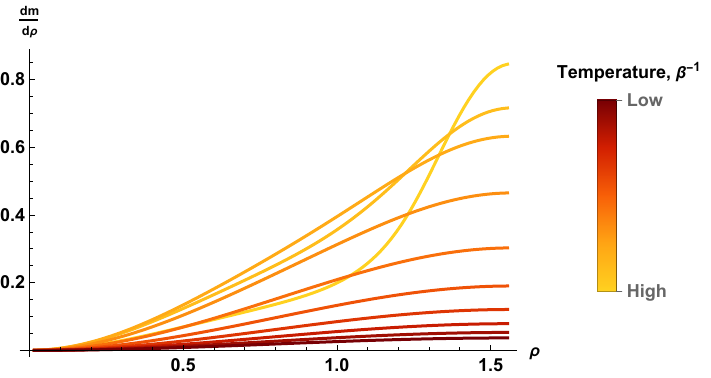}
    \caption{Transparent boundary conditions}
    \label{fig:dmdrhoTransparent}
\end{subfigure}
    \caption{Derivative $\tfrac{\diff{m}}{\diff{\rho }}$ (\ref{eq:meqnrho}) for Dirichlet (top), Neumann (middle) and transparent (bottom) boundary conditions. The inverse temperature lies in the range $\beta  \in (\frac{\pi}{12},\frac{11\pi}{24})$. The inverse AdS length scale is $a=1$. 
    }
    \label{fig:dmdrhoPlotsDNT}
\end{figure}

In this subsection we study the metric function $m(r)$ (\ref{eq:fDefSCEErForm}). 
Its governing equation (\ref{eq:meqn}) is readily integrated numerically, by writing it in the form
\begin{equation}
    \dfrac{\diff{m}}{\diff{\rho }} = 
    - \dfrac{1}{4}  C'(\rho ) {\sqrt {C(\rho ) }} {\widetilde {T}}_{\tau}{}^{\tau}(\rho ),
    \label{eq:meqnrho}
\end{equation}
where all the quantities on the right-hand-side are now functions of the original radial coordinate $\rho $. 
The derivative (\ref{eq:meqnrho}) is shown in Fig.~\ref{fig:dmdrhoPlotsDNT} for Dirichlet, Neumann and transparent boundary conditions.
We see that $\tfrac{\diff{m}}{\diff{\rho }}$ vanishes at the origin $\rho =0$ 
for all three boundary conditions.
Since the energy density $-{\widetilde {T}}_{\tau }{}^{\tau }$ is positive for all $\rho $ (see, for example, Fig.~\ref{fig:RSETplotsD}), it is also the case that $\tfrac{\diff{m}}{\diff{\rho }}$ is positive for all $\rho $ and all boundary conditions.
However, the behaviour of $\tfrac{\diff{m}}{\diff{\rho }}$ as $\rho \rightarrow \tfrac{\pi }{2}$ and the boundary is approached depends strongly on the boundary conditions. 
For Dirichlet boundary conditions (Fig.~\ref{fig:dmdrhoDirichlet}), the derivative $\tfrac{\diff{m}}{\diff{\rho }}$ has a maximum which lies closer to the boundary as the temperature increases, and is monotonically decreasing close to the boundary. 
This is not the case for either Neumann (Fig.~\ref{fig:dmdrhoNeumann}) or transparent (Fig.~\ref{fig:dmdrhoTransparent}) boundary conditions, for which it appears that $\tfrac{\diff{m}}{\diff{\rho }}$ is monotonically increasing towards a finite nonzero value as $\rho \rightarrow \tfrac{\pi }{2}$, and the value on the boundary increases as the temperature increases. 

\begin{figure}
    \centering
    \includegraphics[width=\columnwidth]{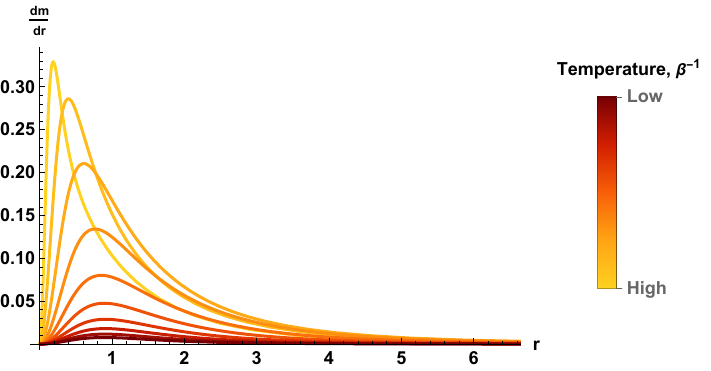}
       \caption{Derivative $\tfrac{\diff{m}}{\diff{r}}$ of the metric function $m(r)$ for Dirichlet boundary conditions. The inverse temperature lies in the range $\beta \in (\frac{\pi}{12},\frac{11\pi}{24})$, and
    the inverse AdS length scale is fixed to be $a=1$.}
    \label{fig:dmdrplotsDN}
    \end{figure}

\begin{figure}
    \centering
    \includegraphics[width=\columnwidth]{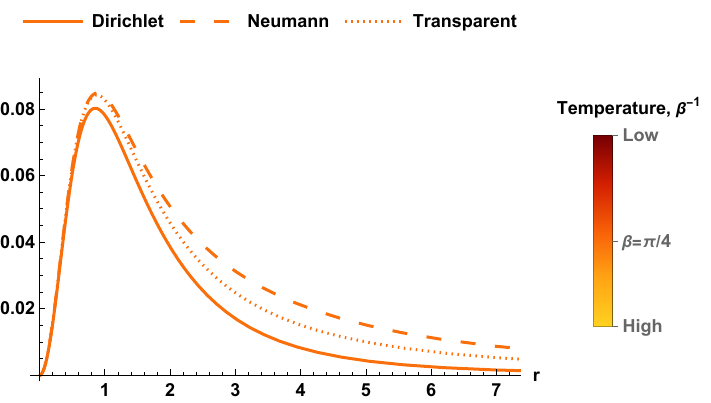}
    \caption{Derivative $\tfrac{\diff{m}}{\diff{r}}$ of the metric function $m(r)$ for Dirichlet, Neumann and transparent boundary conditions at a fixed inverse temperature, $\beta=\frac{\pi}{4}$. The inverse AdS length scale is fixed to be $a=1$.}
    \label{fig:dmdrDirNeuTrans}
\end{figure}

We also consider the derivative $\tfrac{\diff{m}}{\diff{r}}$ (\ref{eq:meqn}), as shown in Figs.~\ref{fig:dmdrplotsDN} and \ref{fig:dmdrDirNeuTrans}.
In Figs.~\ref{fig:dmdrplotsDN} and \ref{fig:dmdrDirNeuTrans} we have evaluated the right-hand-side of (\ref{eq:meqn}) using (\ref{eq:rdef}) and transformed the resulting function of $\rho $ to a function of $r$,  again using (\ref{eq:rdef}). 
In Fig.~\ref{fig:dmdrplotsDN} we present $\tfrac{\diff{m}}{\diff{r}}$ for Dirichlet boundary conditions and a selection of values of the inverse temperature $\beta $.
As expected, $\tfrac{\diff{m}}{\diff{r}}$ is zero when $r=0$, and tends to zero as $r\rightarrow \infty $.
The maximum of $m'(r)$ occurs at a value of $r$ which is temperature-dependent, occurring at smaller values of $r$ for higher temperatures.

Qualitatively similar behaviour is seen in $\tfrac{\diff{m}}{\diff{r}}$ for Neumann and transparent boundary conditions, namely $\tfrac{\diff{m}}{\diff{r}}$ vanishes at $r=0$ and tends to zero as $r\rightarrow \infty $.
We find that the rate at which $\tfrac{\diff{m}}{\diff{r}}\rightarrow 0$ as $r\rightarrow \infty $ depends on the boundary conditions, as can be seen in Fig.~\ref{fig:dmdrDirNeuTrans} for inverse temperature $\beta = \tfrac{\pi }{4}$ (we find similar behaviour for other values of the inverse temperature). 
For larger values of $r$, the derivative $\tfrac{\diff{m}}{\diff{r}}$ takes the smallest values for Dirichlet boundary conditions, and the largest values for Neumann boundary conditions. This is in accordance with the behaviour of $\tfrac{\diff{m}}{\diff{\rho }}$ in Fig.~\ref{fig:dmdrhoPlotsDNT}.

Using the form of the RSET (\ref{eq:RSET}), it is straightforward to show that, for all boundary conditions,
\begin{subequations} 
\label{eq:boundary}
\begin{equation}
    {\widetilde {T}}_{\tau }{}^{\tau } \sim 
    -{\mathcal {T}}_{\infty }^{\lambda }
    \left( \dfrac{\pi }{2} - \rho \right) ^{4} 
    +
    {\mathcal {O}} \left( \dfrac{\pi }{2} - \rho \right) ^{6}
\end{equation}
as $\rho \rightarrow \tfrac{\pi }{2}$, where
\begin{equation}
    {\mathcal {T}}_{\infty }^{\lambda } = \dfrac{a^{2}}{2\pi ^{2}} \sum _{n=1}^{\infty } 
    \left[ n^{3} - \dfrac{\lambda }{9}n \left( 1 + 8n^{2} \right) \right]
    \left( e^{na\beta }-1 \right) ^{-1} .
    \label{eq:Tinf}
\end{equation}
From the plots in Sec.~\ref{sec:ABC}, we see that $C(\rho )/\tan ^{2}\rho $ approaches a finite nonzero limit ${\mathcal {C}}_{\infty }^{\lambda }(\beta )$ as $\rho \rightarrow \infty $, so that 
\begin{equation}
    r^{2}= C(\rho ) \sim {\mathcal {C}}_{\infty }^{\lambda }(\beta )\left( \dfrac{\pi }{2} - \rho \right) ^{-2}
    + {\mathcal {O}} \left( \dfrac{\pi }{2} - \rho \right) ^{0}
\end{equation}
\end{subequations}
as $\rho \rightarrow \tfrac{\pi }{2}$, where ${\mathcal {C}}_{\infty }^{\lambda }(\beta )$ will depend on the boundary conditions applied and the inverse temperature $\beta $.
Combining the results in (\ref{eq:boundary}), we have 
\begin{equation}
   \dfrac{\diff{m}}{\diff {r}} \sim \frac {1}{2} {\mathcal {C}}_{\infty }^{\lambda }(\beta ){\mathcal {T}}_{\infty }^{\lambda} \left( \dfrac{\pi }{2} - \rho \right) ^{2} \sim {\mathcal {O}} (r^{-2})
    \label{eq:mpinfinity}
\end{equation}
as $r\rightarrow \infty $ ($\rho \rightarrow \tfrac{\pi }{2}$).
From (\ref{eq:Tinf}), we have $0<{\mathcal {T}}_{\infty }^{1}<{\mathcal {T}}_{\infty }^{0}<{\mathcal {T}}_{\infty }^{-1}$, which is in agreement with the results in Fig.~\ref{fig:dmdrDirNeuTrans}, since $\lambda $ is given by (\ref{eq:lambda}).

Integrating (\ref{eq:meqnrho}) using {\tt {Mathematica}}'s inbuilt {\tt {NIntegrate}} command gives the metric function $m$ as a function of $\rho$, which can then be transformed to a function of $r$ using the relationship (\ref{eq:rdef}) and the previously computed metric function $C(\rho )$.
For a regular origin at $r=0$, we must have $m(0)=0$, 
which fixes the constant of integration.
Similarly, we find the metric function $\psi (r)$ by integrating the equation for $\tfrac{\diff {\psi }}{\diff {\rho }}$ derived from (\ref{eq:psieqn}):
\begin{equation}
    \dfrac{\diff{\psi }}{\diff {\rho }} = -
    \dfrac{2B(\rho)C(\rho)} {C'(\rho)}
    \left( 
    {\widetilde {T}}_{\tau}{}^{\tau}
    -  {\widetilde {T}}_{\rho}{}^{\rho} 
    \right) .
    \label{eq:dpsidrho}
\end{equation}
We fix $\psi (0)=0$ without loss of generality, since changing the value of $\psi (0)$ corresponds to a rescaling of the time coordinate $\tau $.

For all values of the temperature considered, we integrate (\ref{eq:meqnrho}, \ref{eq:dpsidrho}) from $\rho = \epsilon $ to $\rho = \pi/2-\epsilon $, taking $\epsilon = 10^{-3}$ (changing the value of $\epsilon $ in the range $10^{-3}<\epsilon < 10^{-2}$ results in an estimated relative error of no more than $10^{-6}$ in our results). 
At $\rho = \epsilon $, we use a series expansion for $m$ constructed using the corresponding expansions for the metric function $C(\rho )$ (\ref{eq:Cseries}) and the RSET component ${\widetilde {T}}_{\tau }{}^{\tau }$ (\ref{eq:RSETseries}), which give
\begin{subequations}
 \label{eq:mseries}
\begin{equation}
    m = m_{3} \rho ^{3} + m_{5} \rho ^{5} + {\mathcal {O}} (\rho ^{7}),
\end{equation}
where the $m_{i}$ are constants, the first of which is given by 
\begin{equation}
    m_{3} = -\dfrac{1}{6} C_{2}^{\frac{3}{2}}T_{\tau 0}, 
\end{equation}
\end{subequations}
and the constants $C_{2}$ and $T_{\tau 0}$ are given in (\ref{eq:C2}) and (\ref{eq:Tcoeffs}) respectively.
A similar expansion is used for the metric function $\psi $:
\begin{equation}
    \psi =  \psi _{2} \rho ^{2} + \psi _{4} \rho ^{4} + {\mathcal {O}}(\rho ^{6}),
    \label{eq:psiseries}
\end{equation}
where the $\psi _{i}$ are constants and the lowest order term is given by 
\begin{equation}
    \psi_{2} = -\dfrac{4}{3} C_{2}T_{\tau 0}.
\end{equation}

Using (\ref{eq:Cseries}, \ref{eq:rdef}), the initial value of $\rho = \epsilon  $ corresponds to a very small value $r={\sqrt {C_{2}}}\epsilon $.
The upper limit of our numerical integration, $\rho = \tfrac{\pi }{2}-\epsilon $, corresponds to $r={\sqrt {C(\tfrac{\pi}{2}-\epsilon) }}$. 
As can be seen from Fig.~\ref{fig:DNratioPlotsABCtempvaried}, the value of $C(\rho )/\tan ^{2}\rho $ as $\rho \rightarrow \tfrac{\pi }{2}$ decreases with increasing temperature, and, accordingly, the largest value of $r$ in our numerical evaluation of the metric functions $m(r)$ and $\psi (r)$ also decreases as the temperature increases. 

\begin{figure}
    \centering
    \includegraphics[width=\columnwidth]{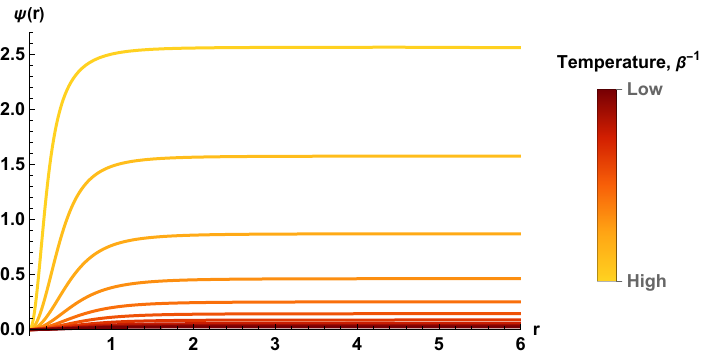}
    \caption{Metric function $\psi (r)$ for transparent boundary conditions.  The inverse temperature lies in the range $\beta \in (\frac{\pi}{12},\frac{11\pi}{24})$ and the inverse AdS length scale is fixed to be $a=1$.}
    \label{fig:psi}
\end{figure}

Our main focus in this section is the metric function $m(r)$, but we first present, in Fig.~\ref{fig:psi}, some results for the metric function $\psi (r)$ when transparent boundary conditions are applied (the corresponding results for Dirichlet and Neumann boundary conditions are qualitatively very similar). 
We see that $\psi (r)$ is a monontonically increasing function of $r$, and that, as $r$ increases, $\psi (r)$ very rapidly converges to a finite limit.
The magnitude of this limit increases with increasing temperature.
The rapid convergence of $\psi (r)$ as $r\rightarrow \infty $ is predicted by the behaviour of $\tfrac{\diff {\psi }}{\diff {r}}$ as $r\rightarrow \infty $, from (\ref{eq:fDefSCEErForm}, \ref{eq:psieqn}, \ref{eq:boundary}):
\begin{equation}
   \dfrac{\diff {\psi }}{\diff {r}}  \sim {\mathcal {O}}(r^{-5}),
\end{equation}
where we have used the fact that ${\widetilde {T}}_{\rho }{}^{\rho }$, like ${\widetilde {T}}_{\tau }{}^{\tau }$, is ${\mathcal {O}}(r^{-4})$ as $r\rightarrow \infty $.

\begin{figure}
    \centering
\begin{subfigure}{\columnwidth}
    \includegraphics[width=\columnwidth]{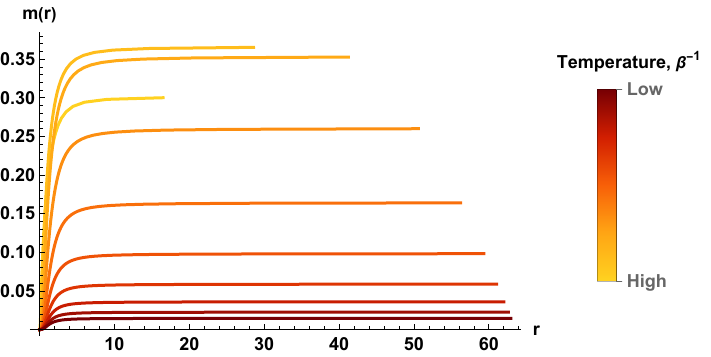}
    \caption{Dirichlet boundary conditions}
    \label{fig:mDirichlet}
\end{subfigure}
\begin{subfigure}{\columnwidth}
    \includegraphics[width=\columnwidth]{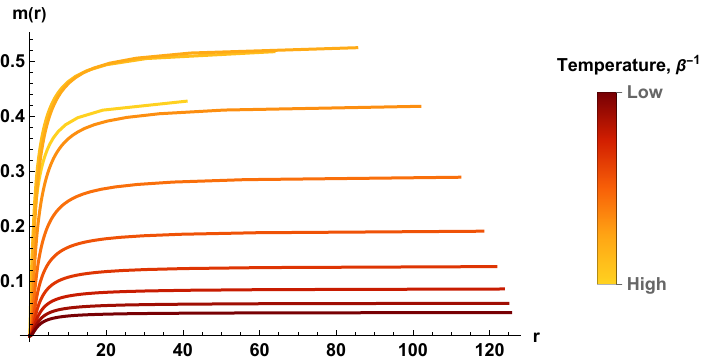}
    \caption{Neumann boundary conditions}
    \label{fig:mNeumann}
\end{subfigure}
\begin{subfigure}{\columnwidth}
    \includegraphics[width=\columnwidth]{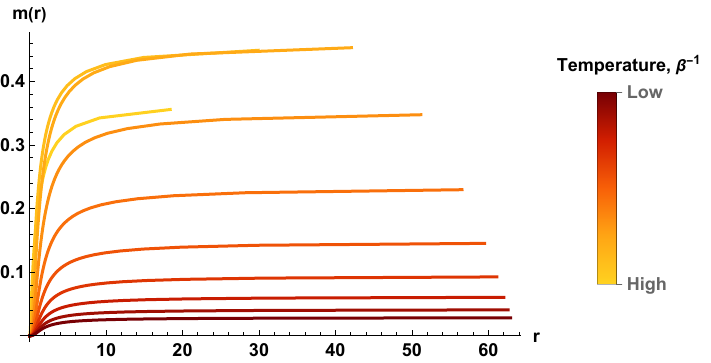}
    \caption{Transparent boundary conditions}
    \label{fig:mTransparent}
\end{subfigure}
    \caption{Metric function $m(r)$ for Dirichlet (top), Neumann (middle) and transparent (bottom) boundary conditions. 
    The inverse temperature lies in the range $\beta \in (\frac{\pi}{12},\frac{11\pi}{24})$.
    The range of values of $r$ shown decreases as the temperature $\beta ^{-1}$ increases. 
    The inverse AdS length scale is fixed to be $a=1$.}
    \label{fig:mrplotsDN}
\end{figure}

We now turn to  Fig.~\ref{fig:mrplotsDN}, where we show the metric function $m(r)$ as a function of the new radial coordinate $r$ when the boundary conditions satisfied by the quantum scalar field are Dirichlet (top), Neumann (middle) or transparent (bottom). 
Since the energy density $-{\widetilde {T}}_{\tau }{}^{\tau }$ is always positive, it follows from (\ref{eq:meqn}) that $m(r)$ is a monotonically increasing function of $r$.
From Fig.~\ref{fig:mrplotsDN}, we also see that $m(r)$ tends to a finite limit as $r\rightarrow \infty $, as expected from the behaviour (\ref{eq:mpinfinity}) of $\tfrac{\diff{m}}{\diff{r}}$ for large $r$.
For fixed temperature, convergence to this limit is quickest when Dirichlet boundary conditions are considered.
Increasing the temperature results in the function $m(r)$ converging more slowly as $r$ increases.

\begin{figure}
    \centering
    \includegraphics[width=\columnwidth]{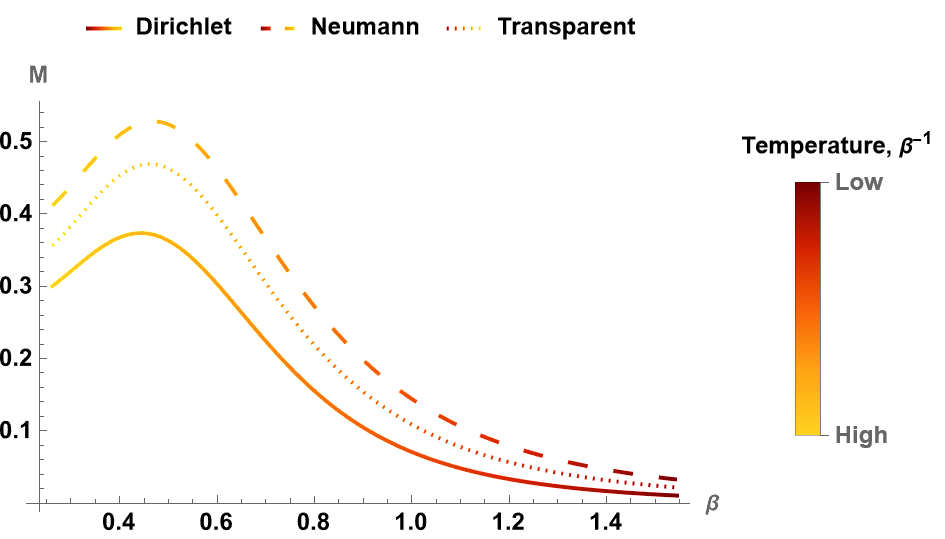}
        \caption{QC soliton mass $M=\lim_{r\to\infty}m(r)$ as a function of the inverse temperature $\beta$, for Dirichlet (solid line), Neumann (dashed line) and transparent (dotted line) boundary conditions.
        The inverse AdS length scale is fixed to be $a=1$.}
    \label{fig:Mtemp}
\end{figure}

The finite limit $M=\lim _{r\rightarrow \infty}m(r)$ is the mass  of the QC solitons. 
To see how this depends on the temperature, in Fig.~\ref{fig:Mtemp} we show $M$ as a function of the inverse temperature $\beta $ for Dirichlet (solid line), Neumann (dashed line) and transparent (dotted line) boundary conditions.
For all fixed values of the temperature considered, we find that the mass of the soliton is smallest when Dirichlet boundary conditions are applied, and greatest when Neumann boundary conditions are applied, with the mass for transparent boundary conditions lying between the Dirichlet and Neumann masses.
At low temperatures (larger values of $\beta $), the mass $M$ increases as the temperature increases ($\beta $ decreases).
However, this trend does not continue for ever-increasing temperature. 
For all boundary conditions, we find a maximum mass $M$ at very similar values of the inverse temperature $\beta \sim 0.5$. 
For larger temperatures, the mass of the QC solitons decreases as the temperature increases further ($\beta $ decreases further).

This result is somewhat unexpected.
For all boundary conditions, the energy density in the quantum field increases with increasing temperature monotonically at all values of $\rho $. 
The total energy in the quantum field can be found by integrating the energy density with respect to $\rho $. 
This gives a finite quantity which is monotonically increasing with temperature for fixed boundary conditions.
Naively one would expect this to be proportional to the mass of the resulting QC soliton. 
As a result, one would anticipate that the QC soliton masses also increase monotonically with increasing temperature, contrary to our results in Fig.~\ref{fig:Mtemp}. 

To understand this,  we write the mass $M$ as follows:
\begin{equation}
    M = -\dfrac{1}{4}\int _{\rho =0}^{\frac{\pi }{2}} 
    C'(\rho ) {\sqrt {C(\rho )}} {\widetilde {T}}_{\tau }{}^{\tau } (\rho ) \, d\rho  .
\end{equation}
From this, we see that the behaviour of the soliton mass $M$ as the temperature increases depends on the behaviour of both the energy density $-{\widetilde {T}}_{\tau }{}^{\tau }$ and the QC AdS metric function $C(\rho )$ with increasing temperature.
While the former (the energy density) monotonically increases with temperature, it can be seen from Fig.~\ref{fig:DNratioPlotsABCtempvaried} that for fixed $\rho $, the QC AdS metric function $C(\rho )$ decreases monotonically with increasing temperature.
The behaviour of the QC soliton mass $M$ as a function of temperature is therefore the result of a complicated interplay between the behaviours of the energy density in the quantum field and the resulting QC AdS metric.

\section{Conclusions}
\label{sec:conc}

We have explored the back-reaction effect of a quantum scalar field on AdS space-time.
We considered a massless, conformally-coupled scalar field on a fixed, global AdS space-time background.
The RSET for this set-up, with the scalar field in either the global AdS vacuum or a global thermal state, can be found in \cite{Allen:1986ty} when the field is subject to either Dirichlet, Neumann or transparent boundary conditions imposed at the space-time boundary.

We used the RSET from \cite{Allen:1986ty} as the source term on the right-hand-side of the QCEE (\ref{eq:QCEE}).
The left-hand-side of the QCEE is purely classical, and describes the QC metric.
When the quantum scalar field is in the global vacuum state, the RSET is a multiple of the metric and the QCEE are trivially solved by a renormalization of the cosmological constant.
In this case the QC metric is simply pure AdS space-time with the inverse AdS radius having a renormalized value.  

To find nontrivial QC metrics, we therefore considered the scalar field to be in a global thermal state, with the right-hand-side of the QCEE being the difference between the thermal and vacuum expectation values of the RSET.
We first solved the QCEE using a metric ansatz (\ref{eq:QCadSansatz}) which facilitates comparison between the QC metric and pure AdS space-time. 
We find that the QC metrics deviate from pure AdS space-time, with the deviations increasing as the temperature of the quantum scalar field state increases.

We then employ an alternative metric ansatz to aid the interpretation of these QC metrics as asymptotically-AdS solitons. 
These solitons have a finite mass which depends on the temperature of the quantum scalar field state. 
For low temperatures, the soliton mass increases as the temperature increases, but it then reaches a maximum. 
The value of the temperature at which the soliton mass has a maximum is very similar for the three boundary conditions studied.
On increasing the temperature further, the mass of the soliton decreases.
This surprising result arises from a complicated interplay between the energy density of the quantum field (which increases as the temperature increases) and the behaviour of one of the QC metric functions (which decreases as the temperature increases). 

We have taken a very simple approach in this paper.
Instead of attempting to solve the full SCEE (\ref{eq:SCEE}), in which the RSET on the right-hand-side is computed on the same background space-time metric which gives the geometric quantities on the left-hand-side, we have fixed the background metric for the RSET, while allowing the QC metric on the left-hand-side to vary (giving the QCEE (\ref{eq:QCEE})).
We have however solved the full nonlinear QCEE, rather than considering linear perturbations about the initial fixed background metric (which, in our case, was pure AdS space-time), following the approach in \cite{Casals:2016odj,Casals:2018ohr,Casals:2019jfo}. 
The QCEE can be considered as an approximation to the SCEE, which will be valid when quantum corrections to the space-time are small. 
In our situation, this corresponds to low temperatures for the quantum scalar field state.
As the temperature of the quantum scalar field increases, we find that quantum corrections to the space-time also increase, and therefore for sufficiently large temperature the approximation used in this work will breakdown. 
A full nonlinear solution of the SCEE would be required to quantitively estimate the temperature at which this breakdown occurs. 
However, one may speculate that this may be at a temperature below that at which the QC solitons we find have a maximum mass. 
If this were the case, then our result that the QC soliton mass is a decreasing function of temperature for sufficiently large temperature may not persist for solutions of the full SCEE.  
Furthermore, one would expect that for sufficiently high temperatures (above the Hawking-Page phase transition \cite{Hawking:1982dh}) a black hole would form.
We are unable to model such a scenario using the methods in this paper (in Sec.~\ref{sec:QCEE} we assume that the QC metric has a regular origin).
We anticipate that the full SCEE would need to be solved to explore this possibility.  

In this work we have considered only a massless, conformally-coupled scalar field and have worked in four space-time dimensions, applying either Dirichlet, Neumann or transparent boundary conditions to the quantum field. 
The advantage of this is that we have been able to use the simple expressions in \cite{Allen:1986ty} for the RSET expectation values.
It would be interesting to explore the properties of the QC metric resulting from a quantum scalar field with different mass or coupling to the scalar curvature, with different boundary conditions applied, or in dimensions other than four.
For example, the back-reaction effect on three-dimensional AdS has been studied in \cite{Belin:2018juv,Belin:2021htw} for the lowest one-particle state, and it would be interesting to extend that analysis to a thermal state.  
Returning to a massless, conformally-coupled scalar field on global AdS in four dimensions, but instead applying Robin boundary conditions, results in a vacuum state which is no longer maximally symmetric \cite{Morley:2020ayr,Morley:2023exv}, and hence we anticipate that the QC metric even for the vacuum state will not be pure AdS. 
Varying the mass, coupling or boundary conditions gives RSET expectation values which are more complicated to compute numerically (see, for example, \cite{Morley:2023exv,Namasivayam}). 
We therefore leave these investigations for future work. 

\begin{acknowledgments}
The work of J.T.~is supported by an EPSRC studentship.
The work of E.W.~is supported by STFC grant number ST/X000621/1.
E.W.~also acknowledges support of the Institut Henri Poincar\'{e} (UAR 839 CNRS-Sorbonne Universit\'{e}), and LabEx CARMIN (ANR-10-LABX-59-01).
This work makes use of the Black Hole Perturbation Toolkit \cite{BHPToolkit}, in particular the {\tt {General Relativity Tensors}} package by Seth Hopper and Barry Wardell was employed for tensor computations.
Data supporting this publication can be freely downloaded from the University of Sheffield Research Data Repository at {\url {https://doi.org/10.15131/shef.data.26023552}}, under the terms of the Creative Commons Attribution (CC--BY) licence. 
\end{acknowledgments}

\bibliography{nonrot}

\end{document}